\documentclass[12pt]{iopart}

\expandafter\let\csname equation*\endcsname\relax
\expandafter\let\csname endequation*\endcsname\relax
\usepackage{amsmath}

\usepackage[T1]{fontenc}
\usepackage{iopams} 
\usepackage{xcolor}
\usepackage{tcolorbox}
\usepackage[shortlabels]{enumitem}
\usepackage{cite}
\usepackage{graphicx}
\usepackage{amsfonts}
\usepackage{amsmath}
\usepackage{booktabs}   % for tables \toprule
\usepackage{multirow}   % multi rows in tables
\usepackage{url}
\usepackage[breaklinks]{hyperref} % for autoref
\usepackage{breakurl}

\usepackage{caption}    % smaller font size for captions
\usepackage{listings}
\usepackage[bottom]{footmisc}
\usepackage[switch, pagewise]{lineno}

\newcommand{\qty}[2]{%
    \ensuremath{#1\,\mathrm{#2}}%
}

\newcommand{\dpm}[1]{\textit{\textcolor{darkgray}{#1}}}

\tcbset{on line, 
        boxsep=1pt, left=0pt,right=0pt,top=0pt,bottom=0pt,
        colframe=white}

\newcommand{\code}[1]{{\fontfamily{cmss}\selectfont\tcbox{#1}}}
\definecolor{deepblue}{rgb}{0,0,0.5}
\definecolor{deepred}{rgb}{0.6,0,0}
\definecolor{deepgreen}{rgb}{0,0.5,0}
\definecolor{lightgray}{rgb}{0.6,0.6,0.6}
\definecolor{codebackground}{rgb}{0.98,0.98,0.98}

\lstdefinelanguage{Toml}{
    comment = [l]{\#},
    morestring = [b]",
}
\begin{document}
\title[Dareplane]{Dareplane: A modular open-source software platform for BCI research with application in closed-loop deep brain stimulation}

\author{Matthias Dold $^{1,2,3}$, Joana Pereira $^{3, 1, 4}$, Bastian Sajonz $^{3}$, Volker A. Coenen $^{3}$, Jordy Thielen $^{1}$, Marcus L.F. Janssen $^{2}$, Michael Tangermann $^{1}$}

\address{$^1$ Data-Driven Neurotechnology Lab, Donders Institute for Brain, Cognition and Behaviour, Radboud University, The Netherlands}
\address{$^2$ Department of Clinical Neurophysiology, Maastricht University Medical Center, The Netherlands;  Mental Health and Neuroscience Research Institute, Maastricht University, The Netherlands}
\address{$^3$ Department of Stereotactic and Functional Neurosurgery, Medical Center – University of Freiburg, Faculty of Medicine, University of Freiburg, Germany}
\address{$^4$ BrainLinks–BrainTools Center, University of Freiburg, Germany}

\ead{matthias.dold@donders.ru.nl}
\vspace{10pt}

\begin{abstract}

\textit{Objective} - This work introduces Dareplane, a modular and broad technology-agnostic open source software platform for brain-computer interface research with an application focus on adaptive deep brain stimulation (aDBS). 
One difficulty for investigating control approaches for aDBS resides with the complex setups required for aDBS experiments, a challenge Dareplane tries to address.

  \textit{Approach} - The key features of the platform are presented and the composition of modules into a full experimental setup is discussed in the context of a Python-based orchestration module. 
The performance of a typical experimental setup on Dareplane for aDBS is evaluated in three benchtop experiments, covering (a) an easy-to-replicate setup using an Arduino microcontroller, (b) a setup with hardware of an implantable pulse generator, and (c) a setup using an established and CE certified external neurostimulator. The full technical feasibility of the platform in the aDBS context is demonstrated in a first closed-loop session with externalized leads on a patient with Parkinson's disease receiving DBS treatment and further in a non-invasive BCI speller application using code-modulated visual evoked responses (c-VEP).

\textit{Main results -} The platform is implemented and open-source accessible on \href{https://github.com/bsdlab/Dareplane}{https://github.com/bsdlab/Dareplane}. Benchtop results show that performance of the platform is sufficient for current aDBS latencies, and the platform could successfully be used in the aDBS experiment. The timing-critical c-VEP speller could be successfully implemented on the platform achieving expected information transfer rates.

\textit{Significance} - The Dareplane platform supports aDBS setups, and more generally the research on neurotechnological systems such as brain-computer interfaces. It provides a modular, technology-agnostic, and easy-to-implement software platform to make experimental setups more resilient and replicable.

\textit{Clinical trial number} - DRKS000287039

\end{abstract}


\submitto{\JNE}

\maketitle
\clearpage

\ioptwocol

\section{Introduction}

A brain-computer interface (BCI) is defined by the BCI Society~\cite{bcisociety2024} as a system that measures brain activity and derives outputs to interact with its environment. This definition also comprises invasive closed-loop systems which are often referred to as brain-machine interfaces (BMI)~\cite{Hofmann2024}.
To realize a BCI, various processing steps and inherent challenges need to be considered: (1) Data is acquired from potentially multimodal sources. Especially for non-invasive signals, this might result in low signal-to-noise ratios (SNR). (2) The neural activity is decoded with algorithmic approaches that can have high complexity. (3) Feedback and control strategies are then applied based on the decoded signal. These strategies often need to cope with non-stationarities in the decoded data distributions. (4) Feedback is provided based on the controller output via hardware interfaces. A potential processing step (5) is the provision of a behavioral paradigm which needs to be accurately synchronized to the data acquisition.
Challenges associated with these five processing steps become even more pronounced in patient experiments, where limited acquisition time is a typical additional constraint.
In the context of deep brain stimulation (DBS) research, and specifically adaptive DBS (aDBS), which motivated this work, domain-specific requirements further include safety considerations for the feedback/stimulation, and performance considerations for feedback latencies.

As a consequence, researchers are investing significant time into the development of software tools to run such bespoke experiments - while on an abstract level, there is a substantial streamlining potential for BCI experiments along the processing steps (1) - (5). 
The proposed data-driven research and evaluation platform for neurotechnology Dareplane is a free, open-source, and modular software solution designed to exploit this streamlining potential.

The following paragraphs introduce DBS and aDBS on a high level to motivate the domain Dareplane is primarily developed for. DBS is typically delivered via a pair of DBS leads each with multiple contacts. When fully implanted, the leads are connected to an implantable pulse generator (IPG) which delivers short electric pulses to the target region, by voltage or current control between individual contacts (bipolar stimulation) or between contacts and the casing of the IPG (monopolar stimulation). Individual pulses are defined by parameters like amplitude, pulse width or shape, and are repeated with a certain stimulation frequency~\cite{Krauss2021}.

DBS modulates neural activity to alleviate symptoms of various neurological conditions, including Parkinson's disease (PD), essential tremor (ET), or dystonia~\cite{Najera2023}. More recently, DBS has also been proposed as a treatment for psychiatric disorders~\cite{Sellers2024, Groppa2023} such as obsessive-compulsive disorder \cite{Wu2020, Coenen2016}, major depressive disorder~\cite{Mayberg2005, Kisely2018, Coenen2019, Alagapan2023}, and Tourette syndrome~\cite{Cagle2022}. Given this wide range of applications, it is not surprising that research on DBS is an increasingly active field~\cite{Harmsen2022}, with multi-center studies and larger participant counts becoming more common in recent publications~\cite{hollunder:2024}. It is important to note that the exact mechanism of action of DBS is still not fully understood, and thus also subject to current research~\cite{Neumann2023, Najera2023, Lozano2019, Hariz2022, Jakobs2019, Hamani2012}. This  emphasizes the need for software platforms supporting this work.

One particular question of interest is whether the therapeutic DBS application, which is applied 24/7 and with fixed stimulation parameters, can be improved by using closed-loop strategies, i.e., realizing aDBS. Such a closed-loop strategy would read out brain activity from the patient, extract an informative marker, and could adjust the stimulation parameters according to a control strategy based on this marker.
In the context of PD, the search for an aDBS strategy became an increasingly popular research topic~\cite{Habets2018} after Little and colleagues~\cite{Little2013} showed that aDBS can achieve superior therapeutic efficacy with only \qty{50}{\%} of the delivered charge of continuous DBS (cDBS). Various other aDBS systems have since been implemented for PD, modulating stimulation amplitude, frequency or pulse width delivered by the IPG~\cite{Arlotti2021, Swann2018, Velisar2019, Pinafuentes2020}. 

Given the more than 10 years of history since the publication of Little et al.~\cite{Little2013}, it is interesting to note that the research in the applied control strategies is not developing in the same pace as the search for informative markers. Control strategies based on thresholds~\cite{Little2013, Arlotti2021, Swann2018, Scangos2021, Roee2021} are the most dominant approach. The few exceptions proposed reinforcement learning~\cite{Gao2023}, proportional(-integral) control~\cite{Pinafuentes2020, Schmidt2023}, or flexible threshold strategies based on time-integrated measures~\cite{Castano2020b}. 

Taking the engineering perspective on aDBS, there is a variety of challenges, which in part applies to other BCI applications as well. First, data acquisition is limited by the accessibility of IPGs~\cite{Krauss2021} with sensing capability for therapeutic aDBS, or the availability of experiments with externalization~\cite{Feldmann2021, Kashanian2021} in the research context exclusively. The recorded data can have a low SNR and is further polluted by physiological and stimulation artifacts~\cite{Rossi2007, Hammer2022}. Decoding the brain state requires considering non-stationarities on multiple time scales~\cite{Tinkhauser2021} and further adapting the stimulation to a patient's motor state~\cite{Busch2024, Roee2021}. On the feedback end, the broad range of stimulation parameters results in a large search space. This makes aDBS control a challenging optimization problem. 

\begin{figure*}[h]
  \centering
\includegraphics[width=\textwidth]{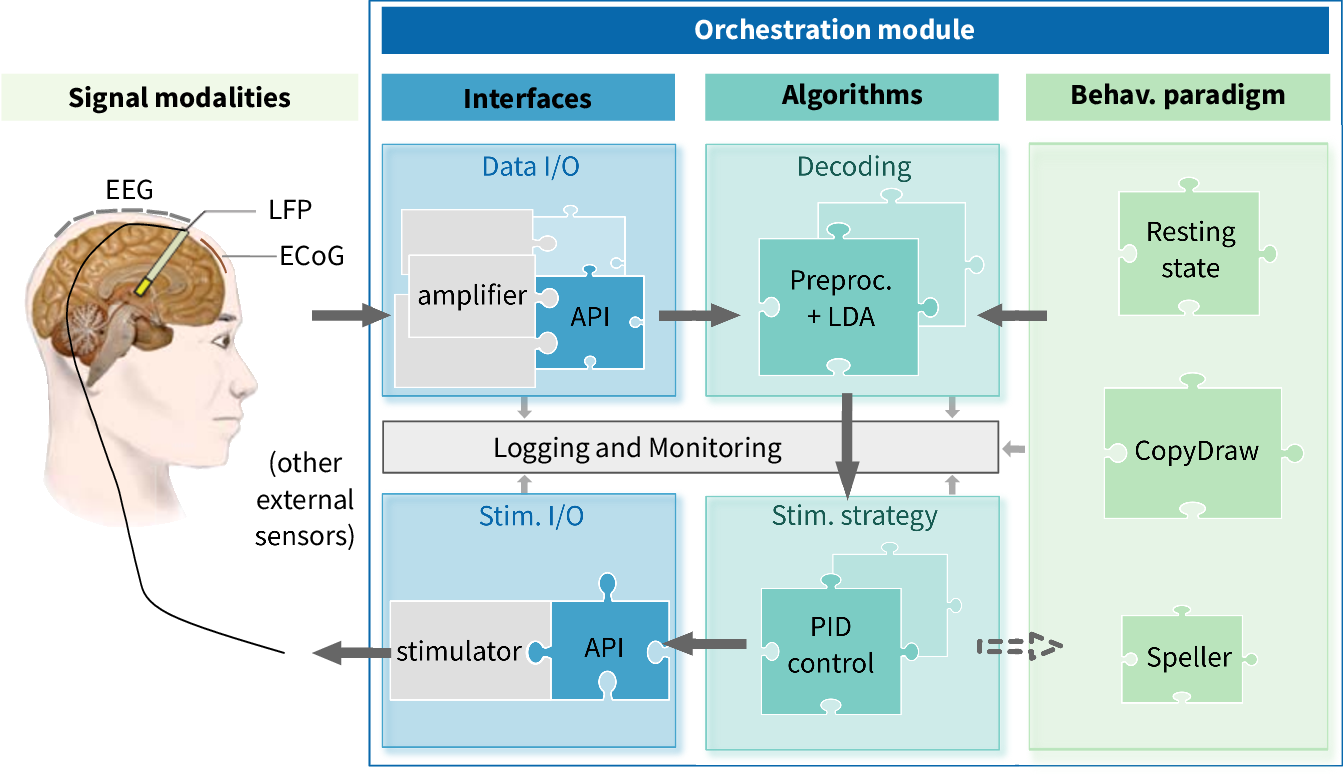}
  \caption{An abstract view of aDBS processing steps. The components can be segmented into three types: 1) I/O for interaction with hardware, either for collecting data or setting stimulation commands (providing feedback for BCIs). 2) Algorithmic components which relate to the decoding of information from the recorded signals and/or evaluate a control strategy based on a biomarker which is extracted from the signal. Such a controller could trigger stimulation in an aDBS setting, or could e.g. select a symbol in a c-VEP speller (see methods section). 3) Presentation of a behavioral paradigm in e.g. displaying a speller matrix or providing a trace to copy by handwriting for the CopyDraw task. Experimenters will usually have a central orchestration of these tasks, to be able to start/stop and modify from a single interaction point. In case the abstract modules - here shown as jigsaw puzzle pieces - are implemented independently, LSL can be used for data transfer between modules and allows a centralized logging and monitoring.}
  \label{fig:dareplane_schema}
\end{figure*}

Putting emphasis on the inherent modular structure, ~\autoref{fig:dareplane_schema} illustrates how the generic processing steps (1) - (5) mentioned above are connected for an example of an aDBS setup. Here,  streaming data from signal recording hardware is realized by a data input/output (I/O) type module. Data preprocessing and decoding, potentially solving a machine learning problem~\cite{Castano2020, Oliveira2023}, are steps of a decoding type module. In the aDBS context, such a decoded signal can be referred to as a biomarker~\cite{Bouthour2019, Neumann2023} if it is also informative about the symtomatology of the disease treated with DBS. Based on the biomarker, a control strategy is then applied to derive a suitable stimulus output by what would be a control module. Such a control module, potentially containing safeguard mechanisms to enforce constraints on its outputs,  would trigger stimulation, or more generally feedback, via what would be a stimulation I/O type module. Finally, a behavioral paradigm, such as the CopyDraw task~\cite{Castano2019}, would be run within a paradigm type module. Changing, e.g., the recording hardware is as easy as replacing the data I/O module in a modular setup, thus keeping the remaining well-tested modules unchanged.

Modularity also provides advantages from a system's design perspective~\cite{Parnas1972, MacCormack2007} and has an extended history in the context of software development for the processing of electrophysiological signals~\cite{Talmon1986}. For many years, modularity has been considered as the key to code reusability~\cite{Gastinger1995} and is encouraged as a decision to be made early in the design process~\cite{Sullivan2001}. It keeps complexity manageable and inherently allows for parallelism in work~\cite{MacCormack2007}. Modularity also leads to well-isolated errors which are faster to fix and thus it facilitates parallel development with multiple programmers. In the academic context, it makes projects more accessible, allowing, e.g., student projects on the limited scope of individual modules.

Looking at the alignment of processing steps in~\autoref{fig:dareplane_schema} it becomes clear that many are generic and could be reused in any open-loop or closed-loop BCI experiment and not just exclusively in aDBS. While, e.g., recording electroencephalography (EEG) signals is a highly generic task, the actual implementation might be equally specific and highly constrained with a vendor's API
only being available from a specific programming language or operating system. A software platform therefore needs to support multiple different hardware interfaces, if reuseability is desired. As a consequence, the platform should be broadly technology-agnostic. 

Finally, aDBS researchers who are not working on a single vendor system exclusively, will use a substantial amount of bespoke software components. To be useful for this target group, and to a large user base in general, any new platform should provide a high level of ease-of-implementation for adopting existing components, minimizing the overhead needed for their integration.

Based on the considerations outlined above, we identify three structural requirements for an aDBS platform: 1) being modular for reusability and robustness, 2) being technology-agnostic to integrate with different hardware and 3) providing ease-of-implementation for integration of existing code/software.

At the point of writing, we are not aware of any existing software platform dedicated to aDBS research. But, as many technical challenges are shared with general BCI applications, would any existing BCI platform meet the requirements for aDBS research? 
Reflecting on a few well-established BCI platforms and some more recent additions, we consider the platforms compiled in~\autoref{tab:platforms}. On a high level, we can categorize the platforms in (a) large scope platforms which partially even come with their dedicated hardware - the “batteries included” package~\cite{Schalk2004, Siegle2017}, (b) platforms which allow users to define computations as graphs~\cite{Renard2010, Ciliberti2017, Ali2024, Clisson2019} with the aim of reducing development complexity, and (c) platforms which predominantly focus on a single programming language~\cite{Memmott2021, Blankertz2016, Venthur2010, Oostenveld2010, Santamaria2023}, requiring implementations of closed-loop setups to follow their coding conventions. The three platform categories address the three requirements (modularity, technology agnostic, and ease of implementation) in different ways. The "batteries included" platforms support many hardware interfaces out of the box, but come with a large core package lacking the ease of implementation for existing code and providing a limited level of modularity. Platforms with graph-based computations are inherently modular in the computation nodes, but require implementations with extended constraints to allow the integration of such nodes. In addition, the technology is typically limited to the main language of the platform (i.e., not technology-agnostic). The single programming language segment by definition is not technology-agnostic and neither is modularity ensured, but it can be realized depending on the actual implementation details.

\begin{table}
    \centering
    \small
    \begin{tabular}{|l|c|c|}
        \hline
        Name & Core Tech. & License \\
        \hline
        \textbf{BCI2000} \cite{Schalk2004} & C++ & GPL\\
        \textbf{Open Ephys} \cite{Siegle2017} & C++ & GPL3\\
        \hline
        \textbf{OpenViBE} \cite{Renard2010} & C++ & AGPL3\\
        \textbf{Falcon} \cite{Ciliberti2017} & C++ & GPL3\\
        \textbf{BRAND} \cite{Ali2024} & C++/Python & MIT\\
        \textbf{Timeflux} \cite{Clisson2019} & Python & MIT \\
        \hline
        \textbf{BCIPy} \cite{Memmott2021} & Python & Hippocr.\\
        BBCI \cite{Blankertz2016} & MATLAB & MIT\\
        PyFF \cite{Venthur2010} & Python & GPL2\\ 
        \textbf{FieldTrip} \cite{Oostenveld2010} & MATLAB & GPL\\
        \textbf{MedusaBCI} \cite{Santamaria2023} & Python & CCPL\\
        \hline
    \end{tabular}
    \caption{Non-exhaustive list of BCI platforms and frameworks for running closed-loop experiments together with the core programming language used/supported and the license which they have been published under. Projects that showed activity since January 2023 are marked in boldface.}
    \label{tab:platforms}
\end{table}

Reflecting back on the processing steps (1) - (5) that need to be fulfilled for an aDBS platform, none of the existing frameworks, except for BCI2000~\cite{Schalk2004}, would provide control of DBS stimulation hardware out of the box. Although open-source software can always be extended, the development effort to add control functionality for stimulation varies strongly between platforms, depending on the imposed constraints and the availability of APIs for the stimulation hardware. As the three requirements (modularity, technology agnostic, and ease of implementation) are only partially fulfilled by existing platforms, and since additional development effort for adding functionality would be necessary, we decided to develop Dareplane as a modular open-source platform to match the requirements and provide functionality tailored to aDBS research.

Like Dareplane, all of the aforementioned platforms are open-source tools. Open-source software in research offers several advantages: it has the possibility to develop features more rapidly, provides more accessible collaboration opportunities, and attracts industry partners for contributions, as evidenced by the success of Lab Streaming Layer (LSL)~\cite{Kothe2024, LSL2024} third-party applications. Having a fully open-source platform also enhances the reproducibility of research results and aligns with open research initiatives common in many fields.
For these reasons, we provide the existing Dareplane modules under the permissive MIT license.

The remaining manuscript follows along four core questions: Q1) How can a modular, technology-agnostic, and easy-to-implement platform be realized? Q2) What latencies for BCI control can be achieved with the proposed platform on different hardware setups, and are they sufficient for aDBS research? Q3) Does the novel platform enable aDBS in a patient experiment with externalized DBS leads? Q4) Is the platform also able to support a code-modulated visual evoked potential (c-VEP) BCI, which servers as an example of a timing-critical application? \\
The questions are answered along the following structure: \autoref{sec:methods} describes the design and capabilities of the platform, reflecting on the relevance for aDBS research, before we present in \autoref{sec:results} the results of performance tests on benchtop setups, results of an aDBS experiment with a patient with PD, and results from three closed-loop c-VEP sessions as proof of technical feasibility. Finally, the platform design and the results are discussed in \autoref{sec:discussion} and concluded in \autoref{sec:conclusion}.

\section{Methods}
\label{sec:methods}

\subsection{Realizing modularity in a technology-agnostic setup}
The following section describes how the Dareplane platform addresses Q1, the question of how to implement a modular, technology-agnostic, and easy-to-implement platform.

Modularity for the Dareplane platform is provided by a segregation of the processing steps and the use of common interfaces. The processing steps (1) - (5), which were introduced above, provide a very high level of separation into single responsibilities. Staying on this high abstraction level has the advantage of limiting the complexity of the interaction protocols of the modules. Alternatively, on the other end of the spectrum from abstract to explicit, modularity could be defined on a class/struct implementation level. This would already prescribe a certain software technology, or require wrapper code, and would require adhering to the naming conventions of attributes and methods. Cutting modules on the fine grained class/struct level leads to constraints, which increase the complexity of the setup, but are irrelevant to the actual processing steps of aDBS. It therefore is helpful to stay abstracted from implementation details and focus on the processing steps that need to be solved, keeping the interaction protocols as unified and unconstrained as possible.
To this end, each module will either consume and/or produce streaming data and will have a limited set of commands exposed for controlling it. 

For sharing data, we decided to make use of the well-established LSL~\cite{Kothe2024, LSL2024} framework which allows the synchronisation of data streams across networks with reliable millisecond precision. LSL is a well-accepted standard with more than one hundred client applications, i.e., software which is freely available and commonly used in the context of BCI research. This includes clients for many recording systems provided by the vendor companies. These clients can be readily used for the data input processing step (1) within the Dareplane platform. LSL can also be considered technology-agnostic as it provides a dynamic library~\code{liblsl}, which has bindings to most programming languages relevant in the BCI field. 

To allow the synthesis of different modules into a complex systems setup, i.e., module interaction, we rely on an API-like structure for each module. In particular, every module communicates via a TCP/IP socket, a basic technology available to almost any relevant programming language. Implementing a TCP/IP communication server can oftentimes be easily wrapped around existing programmes and scripts with little overhead - see the Python example in \autoref{listing:python_wrapper} - addressing the ease-of-implementation requirement. With these minimal requirements on the functionality of a module, Dareplane should be kept attractive for development and extension.

For the communication itself, a simple structure is used which contains a so-called primary command (PCOMM) and potentially an additional json payload. The idea lends itself from the hypertext transfer protocol \cite{RFC9110} (http) methods such as POST or GET. In an abstracted view, the method tells the server how to process the payload which contains key-value pairs relevant for such processing. On Dareplane, every module is free to implement its own primary commands that are then linked internally to the functionality the module wants to expose. A simple \code{'START\_REC'} command would be an example for a data I/O module to start the recording process. A reasonable payload to be sent alongside this command could contain the name of a file to store to, e.g., \code{\{"file\_path": "./data/my\_storage\_file.xdf"\}}.

Requiring modules to expose functionality via TCP communication also allows to integrate modules with other non-Dareplane tech stacks that provide TCP communication. Together with the use of LSL for data streaming, this makes the platform mainly technology-agnostic.

\subsection{Designing a Dareplane module}

Based on the design considerations, any software which can be run in an independent process, exposes its core functionality via the PCOMM structure, and provides and/or produces data to LSL streams, can be considered Dareplane-compliant, i.e., it can be used as a Dareplane module. A comprehensive minimal example can be found in the~\code{hello\_world} example as part of the Dareplane main repository (see \autoref{sec:results}). Additionally, a setup script is provided at \href{https://github.com/thijor/dp-cvep}{https://github.com/thijor/dp-cvep} to download and configure the modules required to run a c-VEP matrix speller. Creating a Dareplane setup with this script provides a wholistic example application, also showcasing how the central orchestration can be used to run experiments.

As most of the current development on the Dareplane platform is done in Python, we provide an utilities library, \dpm{dareplane\_utils}, which can be installed via the \code{pip} package manager as \code{pip install dareplane-utils}.
Getting a custom Python tool wrapped to be Dareplane compliant is then as easy as using the \code{DefaultServer} class from \dpm{dareplane\_utils} and providing it with a dictionary of PCOMMs.
The \code{DefaultServer} is implemented to accept a TCP connection and has a parser which will interpret incoming PCOMM messages to invoke the callable that is matched by key of the prefix. 
For the example provided in \autoref{listing:python_wrapper}, a PCOMM 
message of \code{DECODE|\{"n": "1"\}} would lead to a function call equivalent to 
\code{run\_decode(n="1")} in Python.

The implementation of such a communication layer can then be tested independently by any tool capable of creating a TCP connection and sending text messages, e.g., using \code{telnet}. While the TCP socket allows any binary string to be communicated, there is a limitation depending on vow the message is decoded. For the current Python modules using the \code{DefaultServer} from the dareplane-utils, PCOMMs are decoded as ASCII and therefore need to contain valid ASCII symbols. Furthermore, the pipe symbol (\"|\") is used as a delimiter between the command and the payload and must therefore not be used within either. Finally, the PCOMM is supposed to end on a semicolon (\";\") which is therefore also a reserved symbol.

\begin{lstlisting}[language=Python,caption={Python wrapper example to expose the function \code{run\_decode} as a Dareplane module.},captionpos=b,label={listing:python_wrapper}]
from dareplane_utils.default_server.server import DefaultServer

# Assuming there is a `decoder.py` script containing the callable `run_decode`
from decoder import run_decode 

def main(port: int = 8080, ip: str = "127.0.0.1"):

    # The callable function `foo` will be exposed as PCOMM `DECODE` 
    pcommand_map = {"DECODE": run_decoder}

    server = DefaultServer(port, ip=ip, pcommand_map=pcomm_map, name="decoder_server")
    server.init_server()
    server.start_listening()

\end{lstlisting}

\begin{figure*}[!ht]
  \centering
\includegraphics[width=\textwidth]{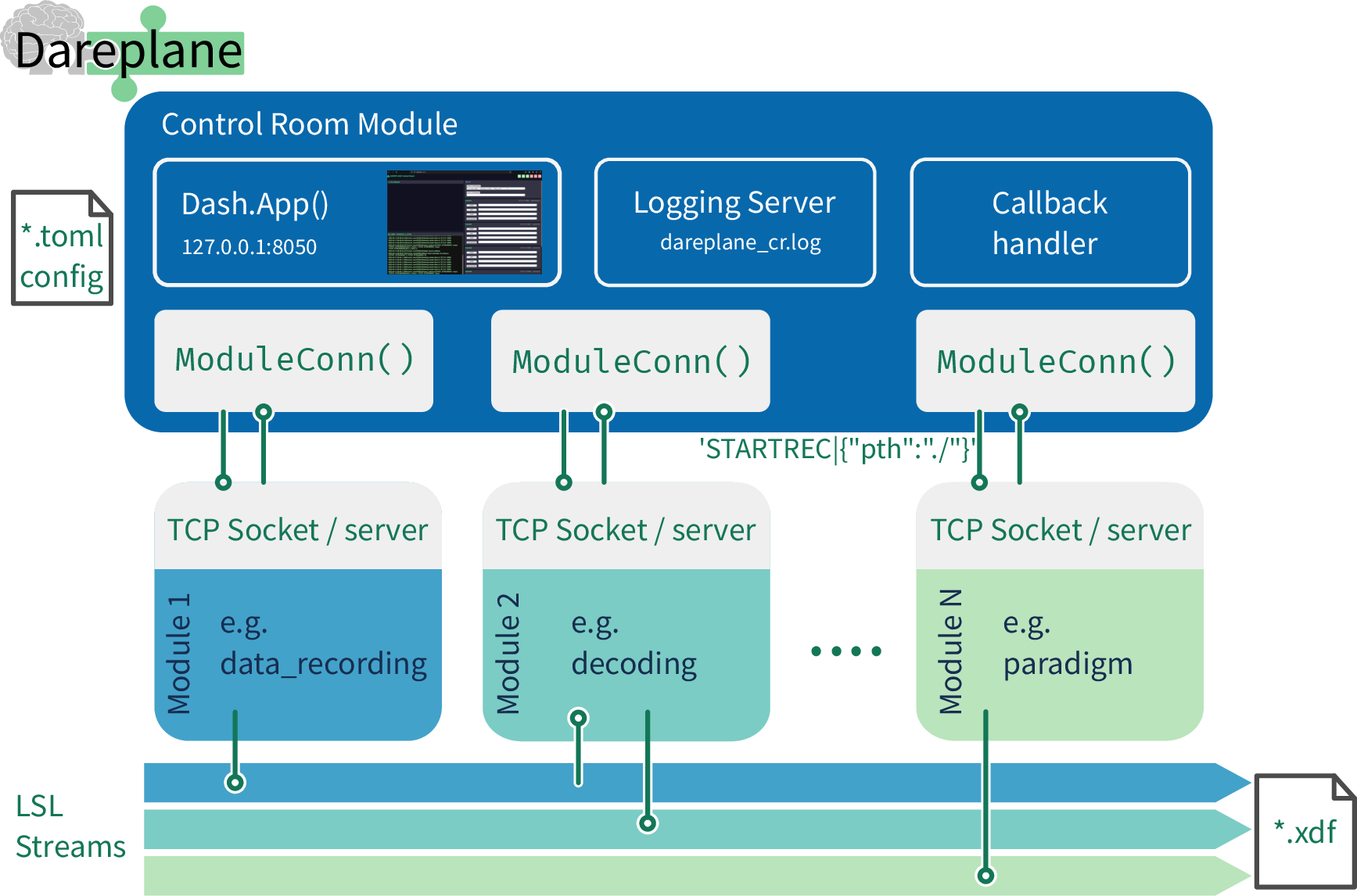}
  \caption{Control room orchestration schema. The control room module orchestrates a Dareplane setup by spawning different modules as subprocesses and connects to them via TCP sockets for sending primary commands such as e.g., \code{STARTREC|\{"pth":"./"\}}. Individual modules communicate data via LSL streams where each module can send to or listen to one or multiple streams. Modules can also trigger a function of another module by sending a callback via TCP/IP which is processed in the control room. The user interaction is done via a web GUI which is served by the control room on the localhost. A central logging server is started at port 9020 which is the default for Python TCP logging, allowing individual modules to log to a single log file.}
  \label{fig:control_room_setup}
\end{figure*}

\subsection{Managing modules with the Dareplane control room}

The control room module (\dpm{dp-control-room}) is used to compose multiple Dareplane modules into a Dareplane setup. It orchestrates other modules which are configured in a TOML file. Having a central orchestration or master module is a well-established pattern for modular designs~\cite{Parnas1972}.

The \dpm{dp-control-room} manages different modules as subprocesses for which it provides the bookkeeping, i.e., managing the life cycle from spawning the processes, it connects via TCP/IP to the module's TCP servers (each running at its own port on the local host), and finally it stops the process and potential child processes during a shutdown. Besides the module connections, the control room provides a web graphical user interface (GUI), a logging server that consolidates logging messages into a single log file, and a callback server which allows PCOMMs to be sent across modules. A schematic overview of this functionality is provided in \autoref{fig:control_room_setup}.

By spawning modules as individual subprocesses, Dareplane setups are inherently asynchronous across different modules. Managing the concurrency will then be a task of the operating system kernel, which is highly optimized for this purpose. Concurrency can be increased even further by designing individual modules to perform their internal tasks concurrently, if necessary.

\begin{lstlisting}[language=Toml,caption={Configuration example of adding a module \dpm{dp-cvep-decoder} to a Dareplane setup. The module will have its communication server available at \code{127.0.0.1:8082}.},captionpos=b,label={listing:module_setup}]
[python]
# path to the root of the modules
modules_root = "../"

# names of a module to be used (folder name)
[python.modules.dp-cvep-decoder]                                      
type = "decoder"
# if no port is provided, a random free port is chosen
port = 8082                                                                     
ip = "127.0.0.1"

\end{lstlisting}

\subsubsection{Configuration of Dareplane setups}
Modules are loaded from the \dpm{dp-control-room} according to TOML configurations, e.g., in a \code{config.toml} file.
An example of such a configuration is provided in \autoref{listing:module_setup}. It specifies that modules based on Python will be located in the \textit{modules\_root} folder. In this folder, it would look for a module~\dpm{dp-mockup-streamer} in a folder with the same name and would start the module's TCP server under \code{127.0.0.1:8082}.

All modules will have individual controls, i.e., buttons for their PCOMMs visible in the web GUI. In case a chain of multiple commands is to be started, \textit{macros} can be defined inside the same \code{config.toml} containing the modules for loading. 
Each macro is defined by a unique name, a chain of commands it executes according to the alphabetical order of the keys they are defined under, and macro parameters which can be passed to the individual processing steps.
\autoref{listing:macro_setup} provides a minimal example which shows how a macro can be used to combine calling the functionality of multiple modules. 
In the example, three steps are executed. First, the \dpm{dp-cvep-decoder} loads the model for the file name provided in the variable \code{fname}.

\begin{figure*}
  \centering
\includegraphics[width=\textwidth]{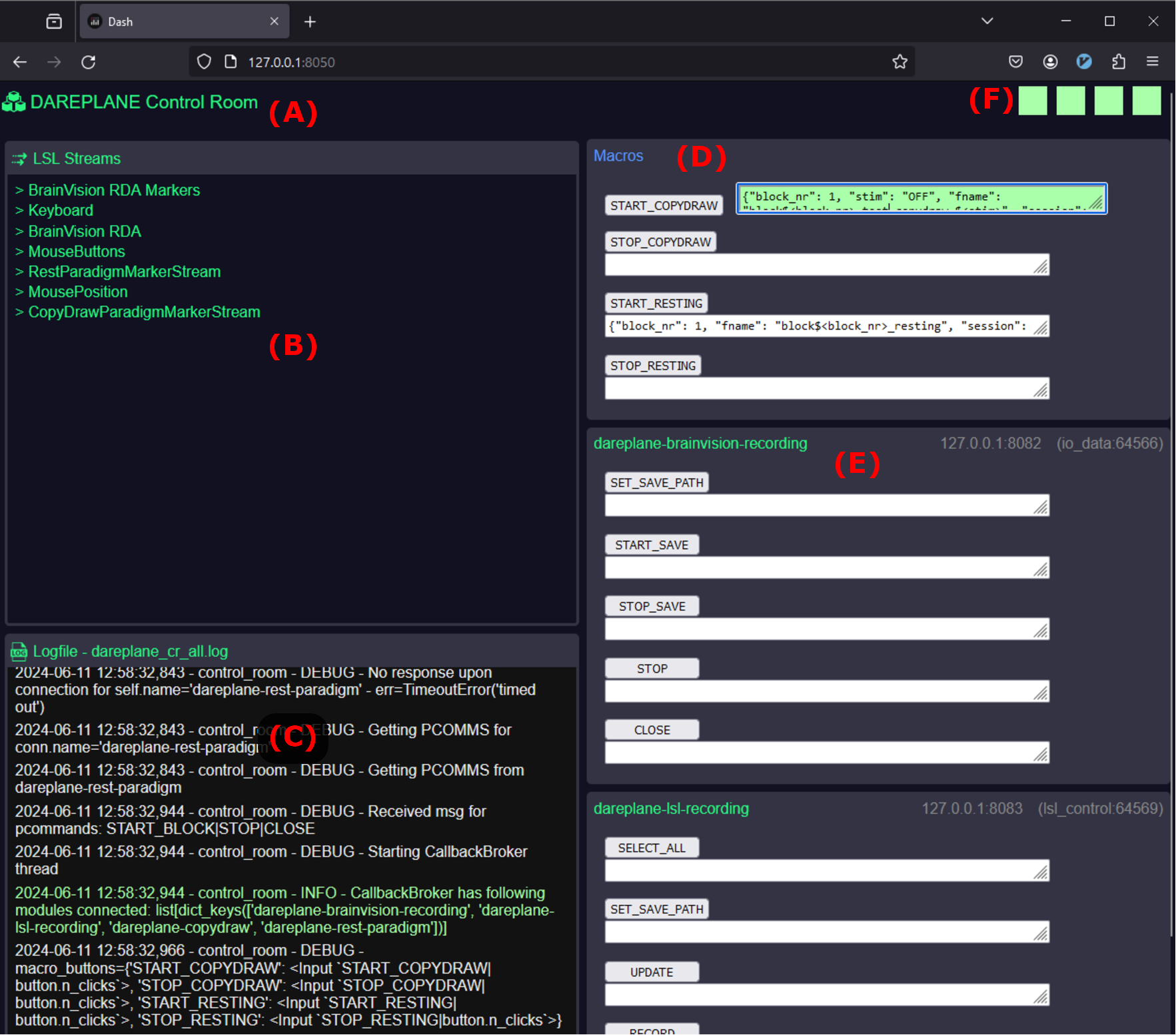}
  \caption{The web application spawned by the \dpm{dp-control-room} module. (A) Shows the application running on localhost at port 8050. It could be reached by other devices depending on the network settings. (B) A tile on the GUI lists active LSL streams. (C) The last entries in the central log file are shown in another tile. Both, (B) and (C), are updated in \qty{3}{s} intervals. (D) The macro section is populated with pairs of buttons for running the macro and a free text field for providing parameters in json format. The content of the free text field is pre-filled according to the `default\_json` values in the config TOML file. (E) A module section is spawned for each configured module, exposing all available primary commands, providing quick access testing and bug fixing. (F) Indicator blocks for each module show if the module server is responding via TCP. The control room is pinging the module servers every \qty{3}{s} and adjusting the color to red if no response is returned. The boxes provide details about the module servers on hover.}
  \label{fig:web_gui}
\end{figure*}

The command is triggered by the \dpm{dp-control-room} sending\\
\code{LOAD|\{"modelfile": "test\_cvep\_1.joblib"\};} to the \dpm{dp-cvep-decoder}.
The payload in this PCOMM message is composed of the key value pairs in the \code{com1} list of the configuration. Additional key value pairs can be provided in additional entries in the configuration list. 
While the key (left side, e.g., \code{modelfile}) will be used in the payload string, the right side refers to variables defined in the \code{default\_json} section of the macro definition. Any key value pairs of this section will appear in the GUI, see \autoref{fig:web_gui} (D). A template structure with \code{$<key\_name>$} is used to conveniently change multiple variables. 
In the provided example, changing the value of block to \code{block=2} in the GUI, would result in using \code{fname$=$"test\_cvep\_2.joblib"} when running the macro. 
Changes of these key value pairs are evaluated once the macro button is pressed.

\begin{lstlisting}[language=Toml,caption={Configuration of a macro for 'start\_online'. First, the decoder module loads a model file, using the \code{LOAD} PCOMM with a payload of \code{\{"model\_file": "test\_cvep\_1.joblib"\}}. Next, the decoder will connect to the data stream by calling the \code{CONNECT} PCOMM. Finally, the \code{RUN\_ONLINE} PCOMM of the \dpm{dp-cvep-speller} module is called. All this can be triggered by a single button press. The button will be labelled with 'START Online'. All PCOMMs need to be implemented in the according modules (exposed from their servers). Macros are defined by the user in config TOML files.},captionpos=b,label={listing:macro_setup}]
[macros]
[macros.start_online]
    # the name will appear on the button in the web GUI
    name = "START Online"
    description = "Start an online run with the speller and the decoder"
[macros.start_online.default_json]
    block = 1
    fname = "test_cvep_$<block>.joblib"
[macros.start_online.cmds]
    # [<target_module>, <PCOMM>, <kwarg_name1 (optional)>, <kwarg_name2 (optional)>]
    com1 = ["dp-cvep-decoder", "LOAD", "modelfile=fname"]
    com2 = ["dp-cvep-decoder", "CONNECT"]
    com4 = ["dp-cvep-speller", "RUN_ONLINE"]
\end{lstlisting}

The second step of the macro will execute the \code{CONNECT} command, e.g., to connect the decoder to the LSL stream of the paradigm.
In this case the PCOMM would not contain a payload and would be \code{CONNECT|;}. The last command of the macro would invoke the \code{RUN\_ONLINE} again with a PCOMM without payload, i.e., \code{RUN\_ONLINE|;}.

Once configured, the \dpm{dp-control-room} creates a web application based on the Python library dash. It runs on the localhost at port 8050, which can be accessed via any web browser, see~\autoref{fig:web_gui} (A).
Having a local web server allows to make the control room accessible to other devices, provided the local network is setup accordingly.
On the webpage (GUI) the experimenter is provided with information about active LSL streams \autoref{fig:web_gui} (B), the latest lines of the central logging file \autoref{fig:web_gui} (C) and the status of the servers for each individual module \autoref{fig:web_gui} (F).
These views update on a \qty{3}{s} interval. \autoref{fig:web_gui} (D) shows a segment which is populated with the macros specified in the config TOML.
For each macro, a button for running the macro and a free text field for providing parameters in json is created.
The content of the free text field is populated according to the `default\_json` values in the config TOML file - c.f. \autoref{listing:macro_setup}.
A validation routine will change the background of the text field if the provided text is not a valid json string.  \autoref{fig:web_gui} (E) shows the sections for each individual module, providing a button and text field for each primary command exposed by the module. These buttons are especially useful for system tests or during debugging as they allow a quick access to all of a module's exposed functionality.

\subsection{Closed-Loop Communication}
Using individual modules spawned up via the control room allows to easily compose complex setups. Up to this point, it was shown how functionality can be made Dareplane-compliant and how modules are managed by the \dpm{dp-control-room}. The final step for a closed-loop setup is to understand how modules interact. Dareplane provides two different approaches for inter-module communication:

\begin{enumerate}[{(1)}]
  \item Using an \textbf{LSL stream} which is populated by the source modules and which is subscribed to by the target modules.
  \item Using \textbf{primary command callbacks} which are triggered via the control room by TCP/IP communication.
\end{enumerate}

Using an LSL stream named, e.g., “decoding2paradigm”,  would be the most canonical approach to allow the sharing of time series data or discrete markers. The target module would be implemented with a listener (LSL inlet) on the given LSL stream and would process the stream values. In this example, the “decoding2paradigm” stream could contain a marker emitted from a decoding source module (LSL outlet), e.g., \code{selected:A} for a BCI speller setup. Then the target module, e.g., the speller paradigm module could stop iterating through the speller matrix and mark the letter~\code{A} as chosen. Analogously, a source module for closed-loop stimulation could be the control module which derived a target stimulation from decoded biomarkers using its control algorithm. The target stimulation amplitude could then be a time series that the control module would broadcast. Based on this time series, the stimulation module would select a stimulation amplitude.

The second possibility, using primary command callbacks, does not allow the sharing of time series signals, but can trigger primary commands of any target module. It uses the existing socket connection between the individual modules and the control room. Such a primary command callback is triggered by a message adhering to the structure \code{\textless target\_module\textgreater |\textless PCOMM\textgreater |}\code{\{potential}\\\code{json payload\};}, which is sent from any module to the control room via TCP socket connection. The first segment in the message informs the control room about which target module should receive the rest of the message via its socket connection with the control room. The message has to contain a valid PCOMM, which needs to be ensured by the implementation in the source module. This method of closing the loop is implemented in Python (since \dpm{dp-control-room} is a Python module) and is especially suited for less time-critical interventions or helpful if the target module cannot implement an LSL inlet. Furthermore, it can be used if an existing module should trigger any functionality which should be separated from the module itself. As an example, a target module could create backups of a file, which is triggered by a source module that has finished recording to that file. 

%Note, that this is a redundancy which might be useful if an existing module is not suited for the task of the target module, or if the functionality of the target module should be triggered by multiple source modules. 

\subsection{Benchtop Tests}
The performance of running closed-loop experiments with the Dareplane platform was tested in a benchtop environment using the arbitrary waveform generator (AWG) of an oscilloscope as a signal source. These tests were conducted to answer Q2, which investigates latencies that the Dareplane platform can achieve with different hardware in an aDBS closed-loop.  

We conducted three different benchtop tests to evaluate the performance of generic, but realistic scenarios for usingDareplane within an aDBS experiment. As the processing requirements of a full pipeline can be highly varying considering, e.g., sampling frequencies and inference frequency for control decisions, the performance tests were conducted with very simplistic processing requirements. Any reported processing latencies are therefore to be understood as capturing especially the overhead of network and memory I/O induced by pulling and pushing from and to different LSL streams, as well as hardware-specific overheads. All tests were run on the same Thinkpad Notebook containing an Intel i7-8750H CPU, 16\,Gb of RAM and an NVIDIA Quadro P1000 GPU. All modules used for testing are available on github at \href{https://github.com/bsdlab/dareplane-paper}{https://github.com/bsdlab/dareplane-paper}.\\

\begin{figure*}
  \centering
\includegraphics[width=\textwidth]{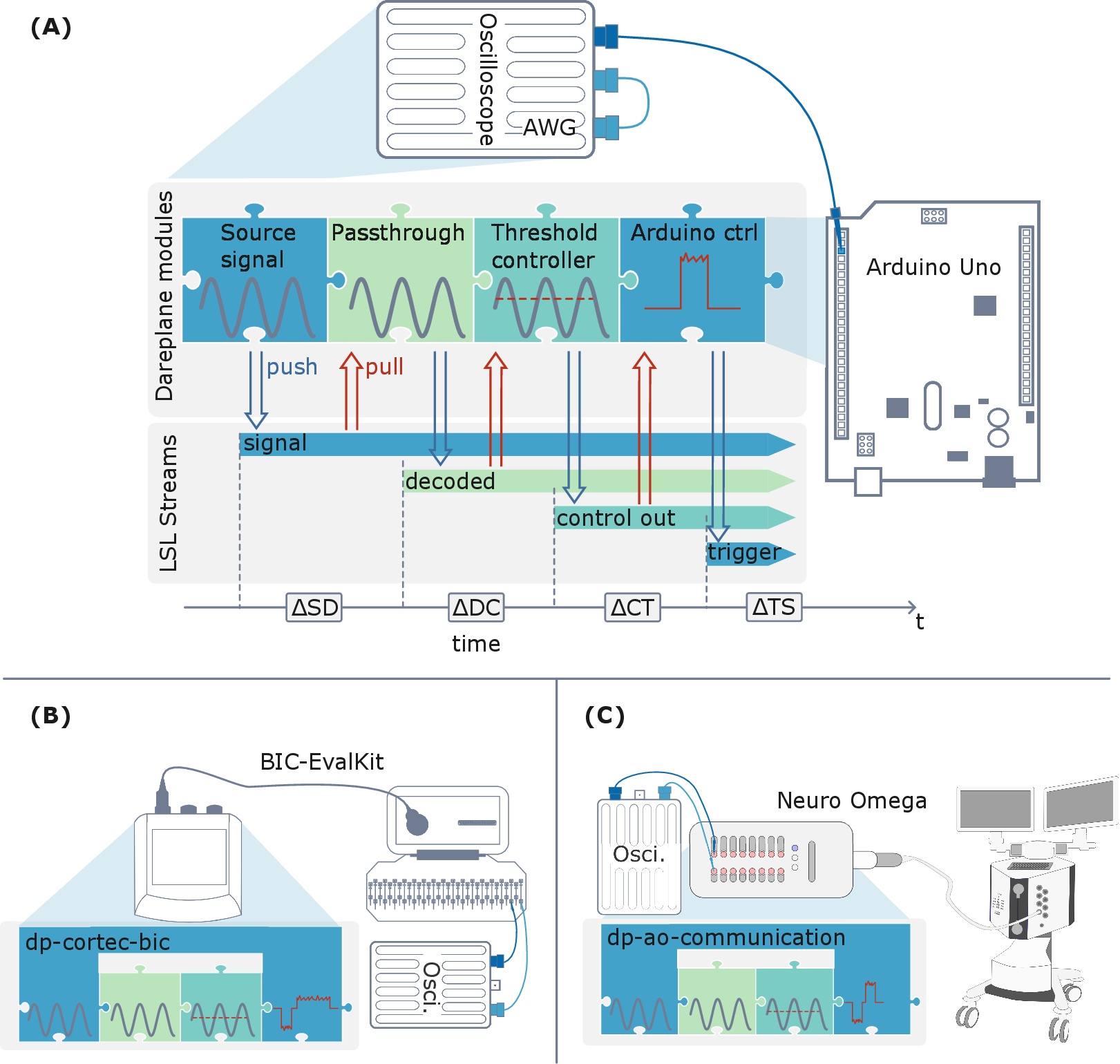}
  \caption{The benchtop setups used for performance testing. \textbf{(A)} shows the setup used for testing with the Arduino Uno as a proxy for a DBS device. The signal input provided by the AWG of an oscilloscope was read via channel 1 from the oscilloscope while the change of the Arduino's GPIO was tracked via a probe connected to channel 2. The "source signal" from the oscilloscope channels was acquired via API by a Dareplane module which streamed it to a "signal" LSL stream.
  This stream was picked up by a "passthrough" module which is in place of a decoding module and passes on data without modification to a "decoded" LSL stream. The time stamps from values of the original "signal" LSL stream and the outbound "decoded" stream were compared to calculate $\Delta$SD. A threshold controller Dareplane module picks up the "decoded" signal and outputs a value of 150 if it is above the threshold and 10 otherwise. The timestamp of the first sample above the threshold in the "decoded" stream was compared to the time stamp of the first sample with a value of 150 in the outbound "control\_out" stream to compute $\Delta$DC. An Arduino control Dareplane module used serial communication to turn the Arduino's GPIO to high whenever a signal of 150 was incoming and turned it to low otherwise. This module is a placeholder for an actual stimulation device API module. The timestamp of the first value of 150 in the "control\_out" stream was compared to the timestamp of the first sample in the "trigger" stream which corresponded to the command of setting the GPIO to high for calculating $\Delta$CT. The final time difference, $\Delta$TS, corresponds to the timestamp difference of the first sample in the "trigger" stream that signalled GPIO high vs.~the occurrence of the stimulation artifact in channel 2 of the "signal" stream as collected from the oscilloscope. \textbf{(B)} shows the BIC-EvalKit setup with a single Dareplane module for signal collection from the BIC-EvalKit's 32 channels, and for sending stimulation commands to the BIC-EvalKit. Both, input and output, had to be combined in a single module as the serial socket only allows for one connection. \textbf{(C)} shows a similar setup as in (B), used for testing with the Neuro Omega. The TCP connection to the Neuro Omega allows only one connection from the client PC, hence again a single module is handling the signal collection and sending of stimulation commands. Other modules and the evaluation of timestamp differences were the same as for (A).}  
  \label{fig:BenchtopTests}
\end{figure*}

The following different hardware components for signal acquisition and the output of stimulation were used:

\begin{enumerate}
  \item \textbf{Arduino} \autoref{fig:BenchtopTests} (A) - This rudimentary setup uses an Arduino UNO (version R3 SMD edition, Arduino S.r.l, Monza, Italy) in place of a DBS stimulator. This setup provides an easy-to-replicate benchmark which does not require specialized and expensive hardware components.

  \item \textbf{BIC-EvalKit} \autoref{fig:BenchtopTests} (B) - This setup uses the Brain Interchange (BIC) EvalKit (version 1.0.2000, CorTec GmbH, Freiburg, Germany), a benchtop system which hardware-wise implements a 1-to-1 copy of CorTec's implant. It was chosen as an example of a system that resembles hardware capabilities of an implantable neurostimulator.

  \item \textbf{Neuro Omega} \autoref{fig:BenchtopTests} (C) - This setup uses the CE certified Neuro Omega (version 1.6.5.0, Alpha Omega, Alpha Omega Engineering, Nof HaGalil, Israel) system. It represents technically mature systems for stimulation via externalized DBS leads. The technical capabilities of this large research-grade system surpass what currently is possible with commercially available DBS implants. It typically is used intraoperatively or on the ward with patients who have externalized DBS leads. 

\end{enumerate}
\paragraph{Raw signal} - All tests used a well-controlled \qty{1}{Hz} sinusoidal source signal with \qty{2}{V} amplitude, which was generated with the arbitrary wave generator (AWG) of a Picoscope 2204A which has an output frequency of up to \qty{100}{kHz}. The generated signal was recorded from either the oscilloscope’s first channel, (Arduino test) or was provided to the stimulation and recording devices by connecting the oscilloscope probe to the recording channels. When the signal was not recorded with the oscilloscope, additional damping elements of $-20$\,dB and $-10$\,dB were used in series to lower the amplitude nominally to \qty{2}{mV}, an admissible range for the CorTec BIC-EvalKit %(\qty{30}{mV_{pp}} for lowest gain) 
and Neuro Omega, not driving amplifiers into saturation. The phase shift induced by attenuation can be neglected, since the signal at the oscilloscope probe (after attenuation) is recorded as the ground truth signal with either BIC-EvalKit or Neuro Omega. Any timing difference is always relative to what these devices recorded.
The signals were read with individual Dareplane modules streaming data from the hardware-specific APIs and pushing the signal to a dedicated LSL stream. Each module was configured to provide data as quickly as possible to the LSL stream. This leads to an output rate which might differ from the sampling rate when the API provides data in chunks. The internal sampling rates were \qty{500}{kHz} for the Picoscope (API configuration with sample interval \qty{400}{ns} and aggregation factor 5) and \qty{5.5}{kHz} for the Neuro Omega. For the BIC-EvalKit a lower sampling rate aligned with the internal \qty{1}{kHz} was chosen to avoid the need for upsampling. While the signal recorded with the Picoscope was a single channel only, the full set of 32 channels for the BIC-EvalKit and 16 channels for the Neuro Omega were streamed to LSL.

\paragraph{Decoding} - The signal was processed in a mock-up decoding module, which performed reading from the signal LSL stream and pushing data back to an LSL stream for decoded data. This mock-up pass-through decoding was chosen since common decoding methods, such as frequency or spatial filtering, can be computed very efficiently - usually within a few microseconds - leaving the bottleneck for processing with the network communication and memory I/O which are represented by the read and write processes. The target sampling rate for LSL output of the decoding module was set to \qty{2}{kHz} for the Arduino and Neuro Omega test. Subsampling for both tests was performed by calculating the median of data received in the \qty{500}{\mu s} intervals. For the BIC-EvalKit, which has \qty{1}{kHz} sampling rate, data was up-sampled by repeating the median of the data buffer (usually one or two samples - see discussion on chunking) to also provide a decoder output with \qty{2}{kHz}.

\paragraph{Control} - Subsequently, a threshold controller picking up the decoded signal, outputting a binary control signal with a numeric value of 150 if the incoming signal was above a predefined threshold or 10 otherwise - the choice of these values is arbitrary. The threshold was chosen to cut the sine wave at approximately half its amplitude of the positive half-wave. This resulted in a threshold at 15000 (equivalent to \qty{1.04}{mV}) for the experiments with the Neuro Omega. In the case of the BIC-EvalKit, which was used with the smallest possible amplification factor of 37.5, and for the Arduino test, a threshold of 100 (equivalent to \qty{100}{\mu V} for the BIC-EvalKit, and \qty{11}{mV} for the Arduino test) was used to stay within the same phase of the sine wave ($[0, \pi/2]$) as for the other tests.

\paragraph{Stimulation trigger} - The controller output was read from LSL by a stimulation module which triggered a single stimulation pulse on the neurostimulators or set the General-Purpose Input Output (GPIO) of the Arduino to high whenever the signal value changed from 10 to 150. \\

The individual Dareplane modules mentioned above resulted in four different LSL data streams which were used to calculate latencies between the individual processing steps.
\newline
$\Delta$SD: time between the raw signal and the output of the pass-through decoder. The time difference was calculated by using the incoming data chunks and their passed through signal. For each chunk, time stamps of the first sample in the signal LSL stream chunk was compared to the first sample of the according chunk in the pass-through stream. Voltage values were compared to ensure correct alignment of samples.
\newline
$\Delta$DC: time between the pass-through signal crossing the threshold and the control signals response. The threshold crossing is calculated as the first data sample above the threshold value. The control signals response time was found by selecting the first output value of 150 following a value of 10. 
\newline
$\Delta$CT: time between the control signal switching to 150 and the control stimulus being sent. The control signal’s switch is determined as for $\Delta$CT, while the time of the control stimulus activation is recorded as a single sample of value different from zero.
\newline
$\Delta$TS: time between the control signal being sent and the stimulation artifact being visible in the raw signal. The trigger time point is calculated as above while the stimulation artifact is identified in the raw signal by applying a threshold value specific to the stimulation hardware\\

All three tests were recorded for more than 5 minutes and cropped during analysis to an exact 5 minute sample window to provide consistent sample sizes across the experiments. 

\subsection{Patient experiment}

Readiness for aDBS, i.e., Q3, was assessed within a series of measurement sessions conducted at the Medical Center - University of Freiburg (Department of Stereotactic and Functional Neurosurgery) on a single patient with PD. The patient (67 years old, female, right-handed) was enrolled in the PD-Interaktiv I study (Extending the Data-Driven Characterization of Neural Markers During Deep Brain Stimulation for Patients with Parkinson’s Disease, DRKS000287039) after providing written consents. This study is ongoing and conducted in accordance with the Declaration of Helsinki and per local statutory requirements. It was approved by the ethics committee of the Medical Center - University of Freiburg (EK application 21-1545). Participants in the study have the DBS system implanted in a two-stage surgical procedure. In the first stage, Boston Vercise\textsuperscript{\texttrademark} (Boston Scientific, Marlborough, USA) directional DBS leads were implanted in the subthalamic nucleus (STN), one lead per STN, and  externalized transcutaneously with special extension wires. An additional four-contact ECoG strip (Ad-Tech Medical Instrument Corporation, Oak Creek, USA) was implanted epidurally over the left primary motor cortex via the same borehole as the DBS lead. The ECoG strip was also externalized. After implantation and externalization, the patient stayed stationary in the ward and took part in the measurement sessions outlined below. In the second surgery, the IPG was implanted, the DBS leads were connected subcutaneously to the IPG, and the ECoG strip was removed. 

The first day after the first surgery (i.e., day 1) was dedicated to recovery. On days two, three and four after this first surgery, measurement sessions, similar to the ones reported in Castaño et al., 2020~\cite{Castano2020}, took place with the following steps: The patient was brought from the ward to the experiment room. The externalized leads for LFP, ECoG electrodes, and four additional electrooculography (EOG) electrodes were connected to the Neuro Omega (Alpha Omega, Alpha Omega Engineering, Nof HaGalil, Israel) system which was used to record signals at \qty{22}{kHz}. Data was streamed using the vendor's C~API and was pushed to an LSL stream for further processing within the Dareplane platform. Additionally, heart rate (HR), respiratory rate (RR) and galvanic skin response (GSR) were recorded on days~2 and~3, using a BrainAmp ExG (Brain Products, Gilching, Germany) system at \qty{5}{kHz} sampling rate. These modalities were omitted on day~4 for a reduced system load during the closed-loop experiment. After connecting the patient, a titration process was conducted with a clinician to establish the electrode contacts and the maximum amplitude to be used for the given measurement day. A frequency of \qty{130}{Hz} with symmetric, initial negative pulses of \qty{60}{\mu s} were used in a monolateral (left or right STN) and bipolar (stimulation and return channel on the DBS lead) stimulation configuration. A bipolar stimulation was chosen as it reduces the amplitude of the stimulation artifact recorded at ECoG electrodes. The stimulation contacts varied between the measurement days and were chosen by the clinician for the subjectively best hand-motor symptom improvement. The configurations during the titration processes are summarized in \autoref{tab:titration}. Please note that during day~2 the stimulation on the left hemisphere was reduced to \qty{4}{mA} during the first CopyDraw recording block as the patient reported a feeling of heat. Once the titration was concluded, the actual measurement sessions started with resting recordings of two minutes each, under stimulation OFF and ON (\qty{130}{Hz}). The resting recordings were not analyzed as part of this work.

\begin{table}
\begin{tabular}{c|r|c|c|c}
\toprule
\multicolumn{2}{l|}{Parameter} & Day 2 &  Day 3 & Day 4 \\
\midrule

\parbox[t]{2mm}{\multirow{3}{*}{\rotatebox[origin=c]{90}{left}}} & stim. contact & 2,3,4 & 5,6,7 & 8\\
 & ret. contact & 5,6,7 & 2,3,4 & 5,6,7 \\
 & amplitude & \qty{6}{mA} & \qty{6}{mA} & \qty{7}{mA} \\
\cline{1-5}
\parbox[t]{2mm}{\multirow{3}{*}{\rotatebox[origin=c]{90}{right}}} & stim. contact & 8 & 8 & 8\\
 & ret. contact & 5,6,7 & 5,6,7 & 5,6,7 \\
 & amplitude & \qty{5.5}{mA} & \qty{6}{mA} & \qty{7}{mA} \\
\bottomrule
\end{tabular}
  \caption{Configuration of the DBS parameters found during the titration processes conducted on every measurement day. Stimulation frequency (\qty{130}{Hz}) and pulse width (\qty{60}{\mu s}) were kept fixed, while a trained clinician selected stimulation and return contacts, as well as amplitude for optimal hand-motor symptom improvement with a bipolar stimulation, separately for the left and right hemisphere. During the CopyDraw experiments on day 2, the stimulation amplitude was reduced to \qty{4}{mA} in the first recording block with stimulation, due to the patient reporting a feeling of heat. On day 4, a stimulation amplitude of \qty{6}{mA} was used to stay compatible with day 3, although higher amplitudes were accepted during the titration process.}
\label{tab:titration}
\end{table}

\begin{figure*}
  \centering
\includegraphics[width=\textwidth]{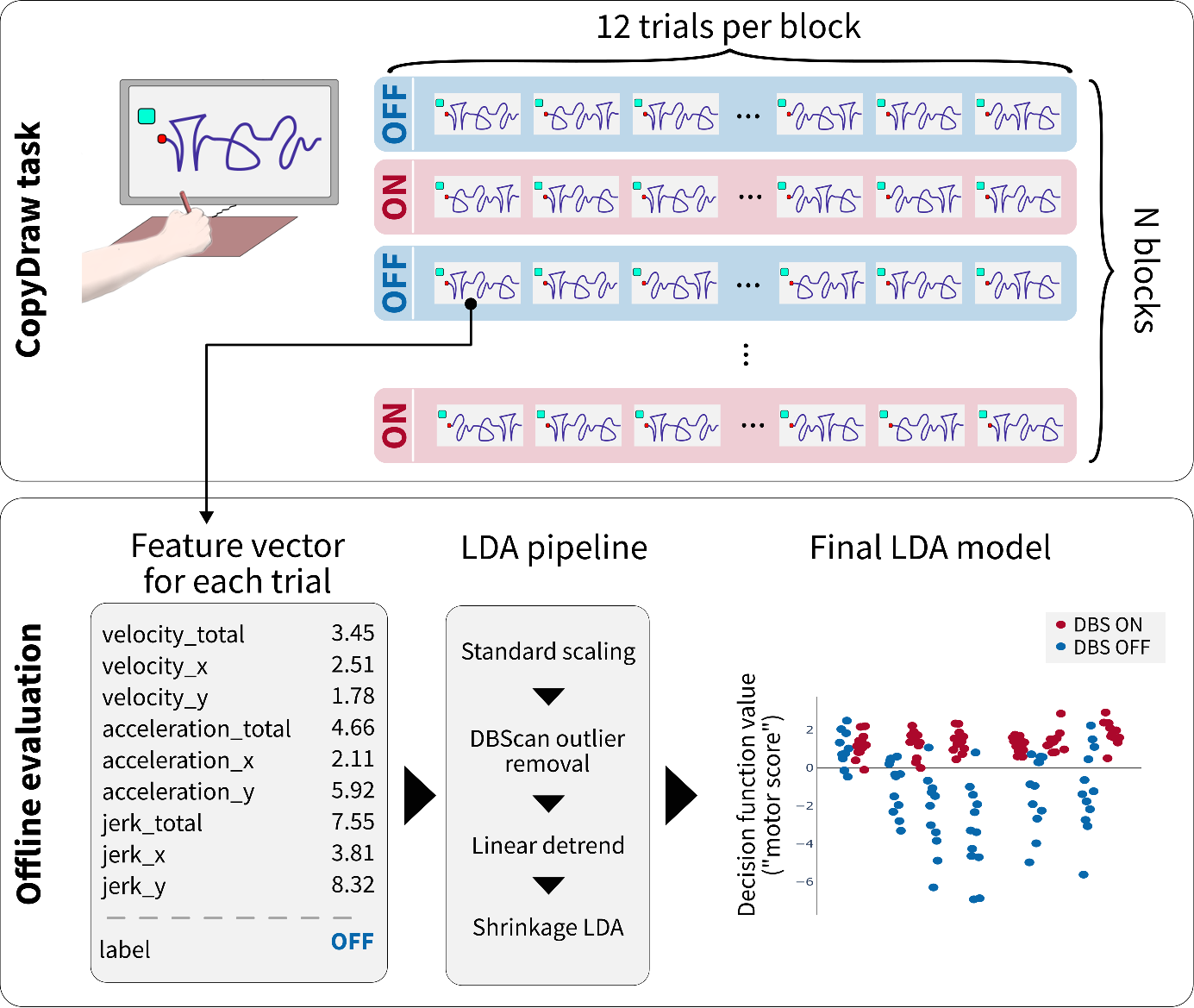}
  \caption{CopyDraw experiment protocol and offline evaluation. The CopyDraw task requires the participant to copy a trace presented on a screen with a stylus on a drawing tablet under time pressure. Twelve trials were collected with a self paced inter-trial-time under a fixed stimulation condition (ON/OFF). Blocks were then collected with iterating stimulation conditions and potential breaks between blocks. After a sufficient number of valid blocks had been collected (usually N=12), an LDA model was trained offline to predict the stimulation condition based on nine behavioral features, containing mean velocity, acceleration and jerk, as Euclidean norm and individual for the different screen coordinate axes. The features were preprocessed by standard scaling, outlier removal using DBSCAN($\epsilon=4$) and removal of a linear trend. An LDA model with Ledoit-Wolf regularization is used to predict the stimulation condition. The decision function values of the LDA model form the CopyDraw motor scores, which serve as a proxy for hand-motor performance capabilities. One such motor score is calculated for each trial. Each point in the scatter plot refers to a single trial.}  
  \label{fig:CopyDrawTask}
\end{figure*}

Next, the main measurements were collected during the CopyDraw~\cite{Castano2019} task \autoref{fig:CopyDrawTask} in stimulation OFF and ON blocks in an alternating pattern. The patient was thoroughly familiarized with the CopyDraw task one day before surgery to avoid a strong learning curve during the first iterations. During each trial (copy-drawing of one trace), the cursor positions were recorded and nine behavioral features per trial were extracted. The features are velocity, acceleration and jerk, each measured in screen x (horizontal) and y (vertical) directions and as Euclidean norm. These features were preprocessed by standard scaling, outlier removal using DBSCAN($\epsilon =4$)~\cite{Ester1996} and the removal of a linear trend by fitting and subtracting a linear regression model for each feature. Next, a behavioral score $y_e \in \mathbb{R}$ for each trial $e$ was calculated as the decision function value of a shrinkage regularized~\cite{Ledoit2004} linear discriminant analysis (LDA) classification model, which had been trained on data of this patient to predict the stimulation state (ON/OFF) with ON being the continuous open-loop stimulation defined by the clinician.
For the purpose of this technical feasibility test, only CopyDraw trials with uninterrupted stylus traces (no pen lift-off during drawing) were considered to ensure comparable behavioral input. A total of 12 blocks with 12 trials each were collected on each recording day. The block structure started with a stimulation OFF block and then alternated between stimulation ON and OFF.

A 6-fold chronological cross validation was applied to gauge decoding performances, assuring that each adjacent ON/OFF block pair was used as the testing fold once. The area under the receiver operating characteristic curve (ROC AUC) was used to measure the decoding performance of the classifier as folds did become imbalanced when only considering traces without a pen lift-off during the trial. See~\cite{Castano2019} for details about the CopyDraw task and~\cite{Castano2020} for an example of how the scores can be used as labels for an EEG decoding pipeline.

\begin{figure*}
  \centering
\includegraphics[width=\textwidth]{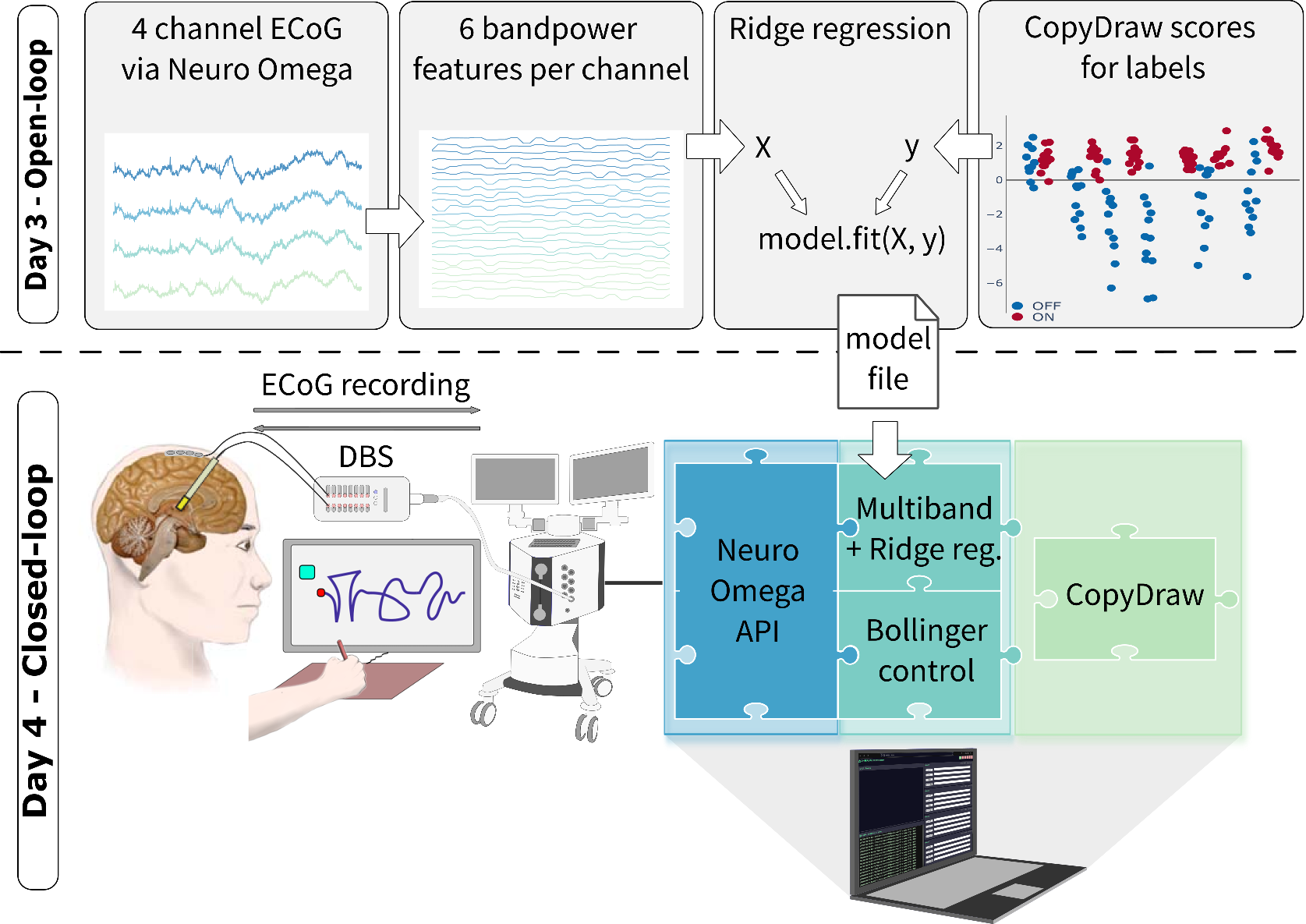}
  \caption{Schematic of the technical feasibility test. The sessions were conducted with a patient with PD who had two DBS leads (implanted to target the right and left STN) and one 4-channel ECoG (over the left motor cortex) externalized.  Open-loop experiments with the CopyDraw tasks were conducted on the second and third days after electrode implantation. The CopyDraw scores and ECoG data from day 3 were used to train a ridge regression model predicting the CopyDraw scores from bandpower features derived from the ECoG data. This model was then used on day 4 together with a Bollinger band control strategy for providing aDBS via the Neuro Omega.}  
  \label{fig:PatientTests}
\end{figure*}

To prove the technical feasibility of aDBS with the Dareplane platform, we trained a decoder to predict the CopyDraw scores based on six pre-defined band power features of the ECoG recordings $\mathbf{x}(t) \in \mathbb{R}^{24}$. The ECoG data was collected with ground placed on the right mastoid and referenced to the left earlobe, while the left mastoid was used as a reference for additional DBS recordings. A linear regression with L2 regularization, a.k.a., a ridge regression model with regularization parameter $\alpha=1$ from the scikit-learn Python library was used for decoding. The features were extracted per ECoG channel using causal 8$^{th}$ order Butterworth band pass filters within the Dareplane decoding module, i.e., using a software filter. The frequency bands were selected as theta (4-\qty{8}{Hz}), alpha (8-\qty{13}{Hz}), lower beta (13-\qty{20}{Hz}), higher beta (20-\qty{30}{Hz}), lower gamma (30-\qty{45}{Hz}) and higher gamma (55-\qty{70}{Hz}). No additional filtering was applied, as the gamma bands did not cover the line noise artifact at \qty{50}{Hz} and no stimulation artifacts were visible in the spectrum of the ECoG channels below \qty{130}{Hz}. After filtering the signal, we estimated the band power by averaging over the absolute values of the filtered signal for a given time window. The average for the offline evaluation, i.e., model training with day 3 data, was chosen across a full trial of CopyDraw of \qty{9}{s}. This trained decoding model was then used on day 4 to provide input to a control module for controlling aDBS - see \autoref{fig:PatientTests}. On day 4, the Dareplane decoding module provided predictions $\hat{y}(t_i)$ of the CopyDraw score at time points $t_i$ with an update rate of either \qty{5}{Hz} or \qty{60}{Hz}. For the online tests (day 4), band power estimates were calculated for each update interval, disregarding the impact of the smaller averaging window especially on the lower bands for this technical proof of concept. The two different update rates were chosen to consider different load stages for the setup, with \qty{60}{Hz} being close to the incoming data chunk frequency that was observed during the benchtop tests with the Neuro Omega - see \autoref{subsec:BenchTopResults}.

The prediction value $\hat{y}(t_i)$ was then used as the input for a control module, which used a Bollinger band strategy~\cite{Bollinger2002} to determine whether to change the stimulation state, motivated by previous work on essential tremor~\cite{Castano2020b}. Bollinger bands are calculated by first computing the moving average of a time series $x\in\mathbb{R}^{n}$ for a given time window of $n$ samples, resulting in $\mu(t_i) = \frac{1}{n} \Sigma_{j = 0}^{n}x(t_{i-j})$. A sample standard deviation $\sigma(t_i) = \sqrt{\frac{1}{n - 1} \Sigma_{j = 0}^{n}(x(t_{i-j}) - \mu(t_i))^2}$  is calculated for the same window. The window length was~\qty{2}{s}.
Bollinger band limits are then defined as $b^{top}(t_i) = \mu(t_i) + \lambda \cdot \sigma(t_i)$ and $b^{bot}(t_i) = \mu(t_i) - \lambda \cdot \sigma(t_i)$ respectively, with $\lambda \in \mathbb{R}$ being a free parameter, which was set to $\lambda =2$. Whenever the upper band limit was crossed ($\hat{y}(t_i) > b^{top}(t_i)$) stimulation was turned off. Whenever the low band limit was crossed ($\hat{y}(t_i) < b^{bot}(t_i)$) stimulation was turned on, which started stimulation with the target amplitude (\qty{6}{mA}) without ramping. To limit the smallest time window in which the stimulation state could change, a \qty{2}{s} grace period was chosen, i.e., within 2\,s of a stimulation change no other such command would be processed. Stimulation commands were sent to the Neuro Omega Dareplane module by means of callbacks. The hyper-parameters were chosen arbitrarily to provide reasonable switching times for this technical feasibility test without any other tuning involved. 

During the patient experiment on day 4, a total of 108 CopyDraw trials were collected under aDBS, with 70 uninterrupted (no pen lift-off) trials. Additionally, 24 stimulation OFF and 36 stimulation ON (continuous stimulation) trials were collected, resulting in 15 and 34 uninterrupted trials respectively (\autoref{fig:CopyDrawDay3}). The model used on day 4 had been trained on all uninterrupted trials from day 3, i.e. N=43 for stim OFF and N=70 for stim ON.

During the offline data analysis, the decoding quality was assessed by the mean Pearson's correlation across the same 6-fold chronological cross-validation as used for the CopyDraw behavioral results. Chance levels were estimated by the \qty{95}{\%} quantile of a bootstrapped permutation distribution (n=2000). Additionally, an ordinary least squares (OLS) regression fit was performed between predicted and true motor scores, using the \textit{statsmodels} Python library. Significance of the OLS fitted parameters was assessed by a t-test for the linear dependency with $\alpha=0.05$ (using the default of statsmodels).

\subsection{BCI c-VEP speller}
\begin{figure*}
  \centering
\includegraphics[width=\textwidth]{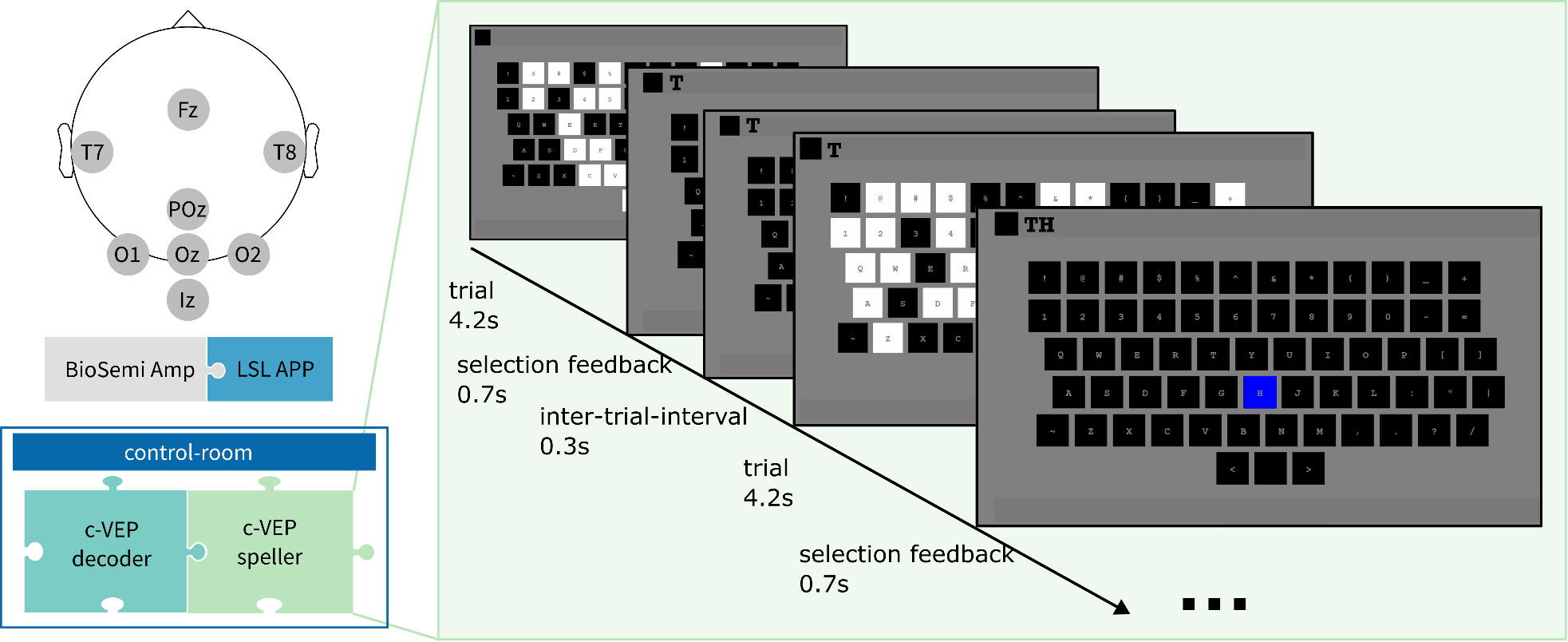}
  \caption{The c-VEP speller setup and paradigm. EEG data is recorded from eight channels using a BioSemi amplifier at \qty{512}{Hz} streamed to LSL using the BioSemi LSL app. Two Dareplane modules are spawned running the speller paradigm and a decoder using rCCA~\cite{Thielen2015} with early-stopping. The paradigm displays a keyboard layout with $n=63$ characters, each flashing in a distinct pseudo-random sequence. Trial periods are up to \qty{4.2}{s}, with the possibility for early stopping. During online use, the decoded character is highlighted for \qty{0.7}{s}, followed by a \qty{0.3}{s} inter-trial-interval.}
  \label{fig:CVEPparadigm}
\end{figure*}

Dareplane also supports setups for classical non-invasive BCIs. As a proof of feasibility, a c-VEP speller, following the setup of Thielen et al.~\cite{Thielen2021}, was implemented on Dareplane to address Q4. A schematic of the setup is shown in \autoref{fig:CVEPparadigm}. Three participants with normal or corrected-to-normal vision participated in the experiments after providing written informed consent. The c-VEP paradigm~\cite{Sutter1992} displays a keyboard-like layout of characters on a screen and simultaneously flashes characters with a unique pseudo-random sequence~\cite{Martinez-Cagigal2021}, as implemented in the \dpm{dp-cvep-speller} module. Each flash sequence evokes a characteristic sequence of brain responses which is captured by eight EEG channels (Fz, T7, T8, POz, O1, Oz, O2, Iz) placed according to the 10-10 system~\cite{Thielen2021}. Data from the training phase is used to train response templates using a reconvolution~\cite{Thielen2015}, which was embedded in a canonical correlation analysis (rCCA)~\cite{Thielen2021} that also learns a spatial filter. The training data included 10 cued trials of \qty{4.2}{s} each, resulting in a total of \qty{42}{s} of training data. The cued targets were selected randomly out of the 63 available characters. 
During the online phase, the calibrated classifier was applied to the growing duration of a single trial every \qty{100}{ms}, which allowed to implement a dynamic stopping approach following\cite{Thielen2021}. For trial reaching the maximum duration of \qty{4.2}{s} a decision was enforced by selecting the character with the highest correlation to a template. 
Upon classification, feedback about the selected character was provided to the user for \qty{0.7}{s}, followed by a \qty{0.3}{s} inter-trial interval. Each participant completed three runs of spelling the target sentence - the quick brown fox jumps over the lazy dog!! - which contains all letters of the alphabet.  All misclassifications had to be corrected by selecting the "<" character, which performed a backspace operation.
The EEG signals were amplified and recorded using a BioSemi ActiveTwo amplifier recording at \qty{512}{Hz} sampling rate. The raw signal was streamed via LSL and processed in the \dpm{dp-cvep-decoder} module. The module continuously bandpass-filtered the EEG data using an 8th-order Butterworth bandpass between \qty{6}{Hz}-\qty{40}{Hz}. This filtered data was then downsampled to \qty{120}{Hz}, twice the refresh rate of the monitor (\qty{60}{Hz}), which allowed classification with the rCCA~\cite{Thielen2015} approach.

\subsection{Scripts and data}
All data and analysis scripts are available at the Radboud Data Repository~\cite{dold2024dataset}.

% %

\section{Results}
\label{sec:results}

\subsection{Dareplane platform}
The Dareplane platform is accessible via its github landing page at: \newline \href{https://github.com/bsdlab/Dareplane}{https://github.com/bsdlab/Dareplane}. \newline Individual modules are referenced therein.
To further enhance accessibility, we provide example modules including a \code{hello\_world} for \href{https://github.com/bsdlab/Dareplane/blob/main/examples/hello\_world/hello\_world.md}{Python} and an \href{https://github.com/bsdlab/dp-c-sdl2-example}{example in~C} using SDL2 for paradigm presentation. For the reproduction of the c-VEP speller setup, a script is provided at \href{https://github.com/thijor/dp-cvep}{https://github.com/thijor/dp-cvep}.

\subsection{Benchtop Results}
\label{subsec:BenchTopResults}

\begin{figure*}
  \centering
\includegraphics[width=\textwidth]{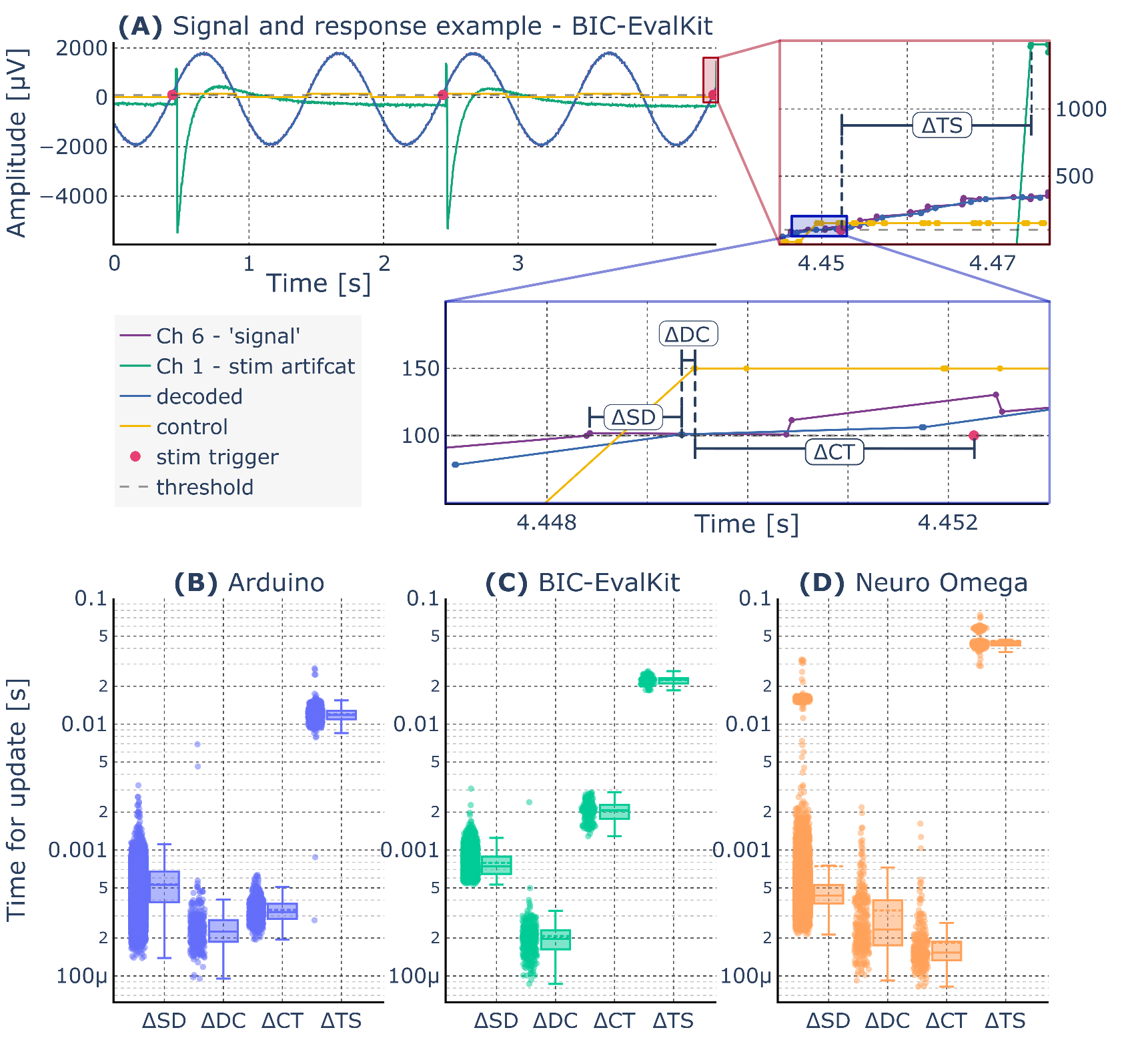}
  \caption{Benchtop performance test results: (A) depicts an example trace of the LSL data streams involved in the BIC-EvalKit testing, showing the \qty{1}{Hz} sinus which is mostly overlapped by the pass-through decoder output. This is an example of how the closed-loop manifest in the recordings, as the crossing of the threshold at \qty{100}{\mu V} off channel six (Ch 6) leads to triggering a single stimulation pulse after passing through the decoding and control modules as visible in the artifact at the channel one (Ch 1). Zoomed in views exemplify how the timing differences were determined. The first zoomed stage (red box) shows how the timing between the stimulation trigger and the stimulation artifact ($\Delta$TS) was determined. The second zoomed stage (blue box) exemplifies how the timing between the incoming signal and the decoder ($\Delta$SD), between the decoder and control output ($\Delta$DC) and between the control output and the stimulation trigger ($\Delta$CT) are determined. The stimulation is triggered by the control modules. (B-D) show box plots of the individual time deltas for the processing steps: $\Delta$SD - signal to decoded, $\Delta$DC - decoded to control response,  $\Delta$CT - control response to trigger being sent, and $\Delta$TS - trigger sent to arrival of the stimulation artifact in the signal.}
  \label{fig:BenchtopPerf}
\end{figure*}

\begin{table*}
\centering
\begin{tabular}{llrrrrrr}
\toprule
 &  & Time [ms] & \multicolumn{5}{r}{} \\
 &  & mean & min & max & median & q1 & q99 \\
Test type & Difference &  &  &  &  &  &  \\
\midrule
\multirow[t]{4}{*}{Arduino Uno} & $\Delta$SD & 0.536 & 0.139 & 3.268 & 0.529 & 0.213 & 0.938 \\
 & $\Delta$DC & 0.277 & 0.095 & 6.917 & 0.225 & 0.119 & 0.543 \\
 & $\Delta$CT & 0.335 & 0.194 & 0.636 & 0.322 & 0.218 & 0.578 \\
 & $\Delta$TS & 12.021 & 0.277 & 27.736 & 11.796 & 8.881 & 16.456 \\
\cline{1-8}
%\multirow[t]{4}{*}{BIC-EvalKit} & $\Delta$SD & 1.051 & 0.553 & 7.890 & 1.041 & 0.597 & 2.117 \\
 %& $\Delta$DC & 0.224 & 0.107 & 0.830 & 0.207 & 0.114 & 0.528 \\
 %& $\Delta$CT & 0.817 & 0.615 & 2.174 & 0.758 & 0.629 & 1.619 \\
 %& $\Delta$TS & 23.342 & 19.655 & 27.193 & 23.260 & 20.081 & 27.162 \\
 \multirow[t]{4}{*}{CorTec EvalKit} & $\Delta$SD & 0.793 & 0.532 & 3.075 & 0.741 & 0.569 & 1.420 \\
 & $\Delta$DC & 0.208 & 0.086 & 2.397 & 0.197 & 0.101 & 0.379 \\
 & $\Delta$CT & 2.055 & 1.289 & 2.871 & 2.072 & 1.394 & 2.819 \\
 & $\Delta$TS & 22.339 & 18.543 & 26.366 & 22.553 & 18.612 & 25.861 \\
\cline{1-8}
\multirow[t]{4}{*}{Neuro Omega} & $\Delta$SD & 0.742 & 0.213 & 32.546 & 0.434 & 0.273 & 15.859 \\
 & $\Delta$DC & 0.331 & 0.092 & 2.188 & 0.234 & 0.105 & 1.494 \\
 & $\Delta$CT & 0.183 & 0.082 & 1.621 & 0.153 & 0.088 & 0.587 \\
 & $\Delta$TS & 45.961 & 28.860 & 73.765 & 43.589 & 32.707 & 61.458 \\
\cline{1-8}
\bottomrule
\end{tabular}
  \caption{Performance comparison of benchtop system tests, showing mean, min, max, median, \qty{1}{\%} and \qty{99}{\%} quantiles. Time differences are: $\Delta$SD - signal to decoded, $\Delta$DC - decoded to control response,  $\Delta$CT - control response to trigger being sent, and $\Delta$TS - trigger sent to arrival of the stimulation artifact in the signal.}
\label{tab:performance_comparison}
\end{table*}

\paragraph{Arduino test:}
This relatively simple test revealed the main bottleneck: The control of the GPIO pin showed a mean latency of \qty{12.021}{ms}, c.f. \autoref{tab:performance_comparison}.
Other modules which are pulling from LSL~\cite{Kothe2024, LSL2024} and pushing to another LSL stream add latency overheads of less than \qty{1}{ms} on average ($\Delta$SD=\qty{0.536}{ms}, $\Delta$DC=\qty{0.277}{ms}, $\Delta$CT=\qty{0.335}{ms}). The value of the \qty{99}{\%} quantile suggest that this performance is overall stable ($\Delta$SD=\qty{0.938}{ms}, $\Delta$DC=\qty{0.543}{ms}, $\Delta$CT=\qty{0.578}{ms}).

\paragraph{BIC-EvalKit test:}
Similar to the Arduino test case, the bottleneck with the BIC-EvalKit was found to be the time between sending out the stimulation command and the appearance of the first stimulation pulse in the signal, with a mean latency of $\Delta$TS=\qty{22.339}{ms}, c.f. \autoref{tab:performance_comparison}. The other modules showed average latencies of $\Delta$SD=\qty{0.793}{ms}, $\Delta$DC=\qty{0.208}{ms}, $\Delta$CT=\qty{2.055}{ms} on average. The example traces in \autoref{fig:BenchtopPerf} show an artifact in the sine wave when readying the system for the next stimulation. The artifact was visible in the whole recording, but was not investigated further as it did not impact the timing measurements.

\paragraph{Neuro Omega test:}
Again, the bottleneck of the closed-loop response is the time between sending out the stimulation command and the appearance of the first stimulation pulse in the signal. We observed a mean latency of $\Delta$TS=\qty{45.961}{ms}.
Consistent with the Arduino tests, the other modules showed average latencies lower than \qty{1}{ms} ($\Delta$SD=\qty{0.742}{ms}, $\Delta$DC=\qty{0.331}{ms}, $\Delta$CT=\qty{0.183}{ms}), c.f. \autoref{tab:performance_comparison}.

\begin{figure*}[!h]
\centering  
\includegraphics[width=\textwidth]{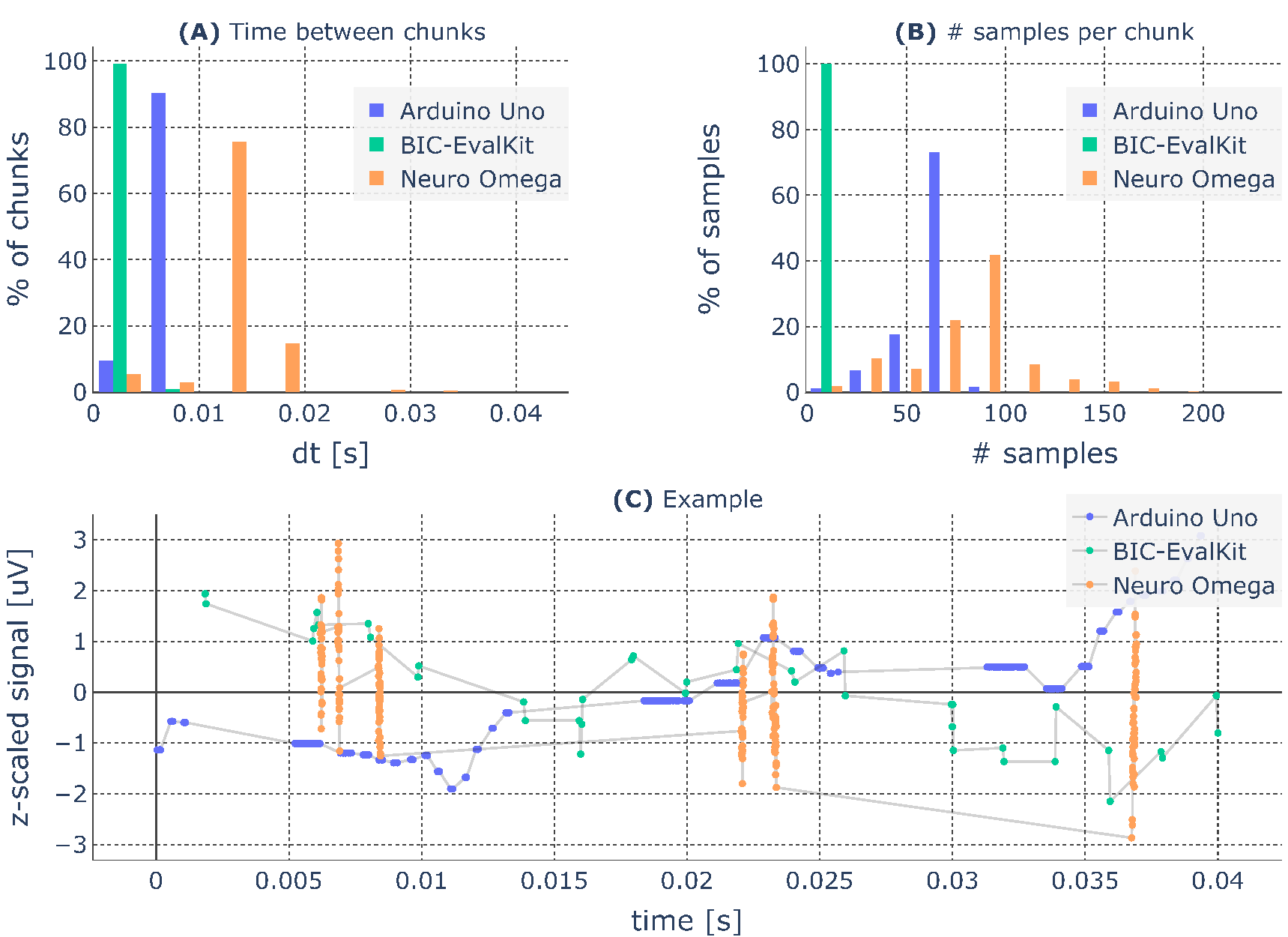}
  \caption{Analysis of chunked data. (A) shows histograms of the time between consecutive data chunks. Chunks were identified by gaps between consecutive sample time stamps >\qty{1}{ms}. Chunks arrive on average with \qty{13 \pm 4}{ms} for Neuro Omega, \qty{7 \pm 2}{ms} for Picoscope, and with almost no chunking (\qty{2 \pm 1}{ms}) for the BIC-EvalKit. (B) shows histograms of the number of samples within each chunk. (C) shows sample traces from each experiment highlighting how an actual realtime processing of the signal would be impacted by chunking. The samples further visualize that the data chunks themselves are incoming in irregular intervals. Note that the Neuro Omega recorded at \qty{5.5}{kHz} sampling frequency, resulting in about 80 samples per chunk being processed.}
  \label{fig:chunks}
\end{figure*}

\begin{table*}[!h]
\centering
\begin{tabular}{lllrrrrrr}
\toprule
   measure & src & mean & std & min & max & q01 & q99 \\
\midrule
dt\_chunk [s] & Arduino Uno & 0.007 & 0.002 & 0.001 & 0.011 & 0.001 & 0.009 \\
dt\_chunk [s] & BIC-EvalKit & 0.002 & 0.001 & 0.001 & 0.024 & 0.002 & 0.004 \\
dt\_chunk [s] & Neuro Omega & 0.013 & 0.004 & 0.001 & 0.045 & 0.001 & 0.029 \\
sample count [\#] & Arduino Uno & 62.364 & 12.282 & 1.000 & 98.000 & 19.000 & 81.000 \\
sample count [\#] & BIC-EvalKit & 2.313 & 0.742 & 1.000 & 6.000 & 2.000 & 4.000 \\
sample count [\#] & Neuro Omega & 79.697 & 29.875 & 1.000 & 226.000 & 1.000 & 163.940 \\
\bottomrule
\end{tabular}
\caption{Chunking statistics of data recorded in the three benchtop experiments. Rows with measure dt\_chunk [s] show aggregates for the time delta between chunks (last to first sample). Rows with measure sample count [\#] show aggregates of the number of samples per chunk.}
\label{tab:chunk_comparison}   % always have label after caption!
\end{table*}

\begin{figure}
  \centering
\includegraphics[width=0.38\textwidth]{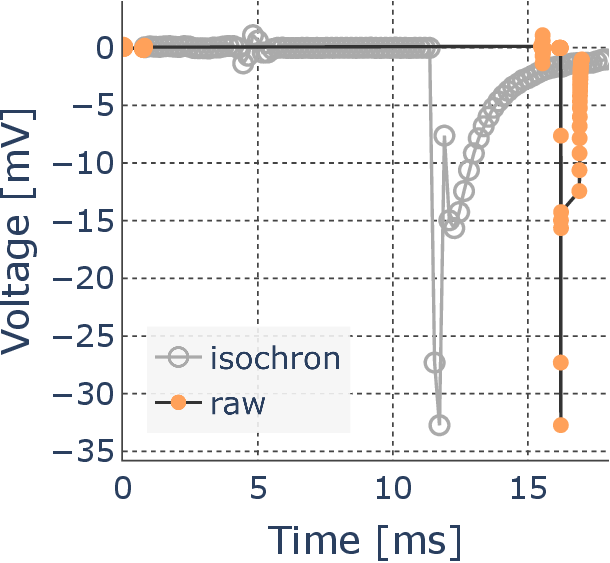}
  \caption{Example of estimating the de-jittered timing of one stimulation pulse with an isochronous approximation. The raw data (orange dots and black trace) recorded from the Neuro Omega reveals a chunking of  \qty{\approx15}{ms}, (see samples up to \qty{1}{ms} vs.~samples from \qty{15}{ms} onwards). Assuming that chunking is caused by data transmission and that the actual samples are recorded with a stable sampling rate, we can estimate the samples of the chunk with time t > \qty{15}{ms} to lie on an isochronous grid (i.e., with regular time distances between sample points) following the last sample of the chunk with t < \qty{1}{ms}. In this isochronous approximation (gray dots and trace), the maximum value of the stimulation artifact is reached at \qty{11.73}{ms}, compared to \qty{16.22}{ms} in the raw signal.}
  \label{fig:AoIsochronAdjustment}
\end{figure}

\paragraph{Influence of chunking}
Although the \dpm{dp-ao-communication} module was requesting data from the Neuro Omega SDK  with a nominal interval of $\approx$\qty{500}{\mu s} (using a sleep command), analysis of the LSL stream showed that even with the sampling rate of $f_{sample}=\qty{5500}{Hz}$ being stable, data is incoming in chunks with a mean inter-chunk time of \qty{13}{ms} [+/- \qty{4}{ms} STD] (\autoref{fig:chunks}). Chunking affected all streamed sources \autoref{tab:chunk_comparison}.
The least chunking was observed for the BIC-EvalKit experiment with an average inter-chunk interval of \qty{2\pm1}{ms} which almost met the native \qty{1}{kHz} sampling rate, followed by the Picoscope (Arduino Uno test) with \qty{7 \pm 2}{ms}, see \autoref{fig:chunks} and \autoref{tab:chunk_comparison}. 
This chunking has a consequence for various measurements. If, e.g., the time-stamp of the stimulation artifact within such a chunk is considered for the calculation of $\Delta$TS, the latency is overestimated by at least the inter-chunk time to the last sample of the previous data chunk.
Considering that the relevant time for aDBS is the arrival of the stimulation pulse at the neural tissue ($t_{stim}$) and not the time point when the artifact shows up in the recorded data ($t_{rec}$), we also calculated the time from sending the stimulation command ($t_{cmd}$) to a de-jittered stimulation time. For this purpose, we created an isochronous approximation $\hat{t}_{stim}\approx t_{stim}$ of the time when the stimulation pulse was actually delivered. \autoref{fig:AoIsochronAdjustment} illustrates this approximation.
Assume the recorded artifact $t_{rec}^{i,j}$ appears at position $j$ in data chunk $i$ (e.g., raw samples with t > \qty{15}{ms} in \autoref{fig:AoIsochronAdjustment}) - the chunk containing a total of $n$ samples. Let $\lbrack t^{i, 1},...,t^{i,j}_{rec},...,t^{i, n} \rbrack$ be the associated time stamps for all samples in that chunk. Using $t^{i-1,m}$, the last time stamp of the previous chunk $i-1$ (e.g., raw samples with t < \qty{1}{ms} in \autoref{fig:AoIsochronAdjustment}) having $m$ samples, the extrapolated arrival time can then be approximated as $\hat{t}_{stim}^i=t^{i-1,m} + j/f_{sample}$ (under the assumption of a stable sampling rate within the amplifier).

Using the corrected stimulation time points $\hat{t}_{stim}^i$ leads to an improved estimate of $\tilde{\Delta}$TS: Mean latencies are reduced from \qty{46}{ms} on raw LSL samples (using $t_{rec}^i$), to \qty{41}{ms} with the de-jittered time stamps. Such a de-jittered estimation was calculated for the Neuro Omega tests only, as almost no chunking was found for the BIC-EvalKit test, and as the Arduino test is just a proxy domain without the claim of replicating stimulation pulses accurately.

\subsection{Patient experiment}
\begin{figure*}
  \centering
\includegraphics[width=\textwidth]{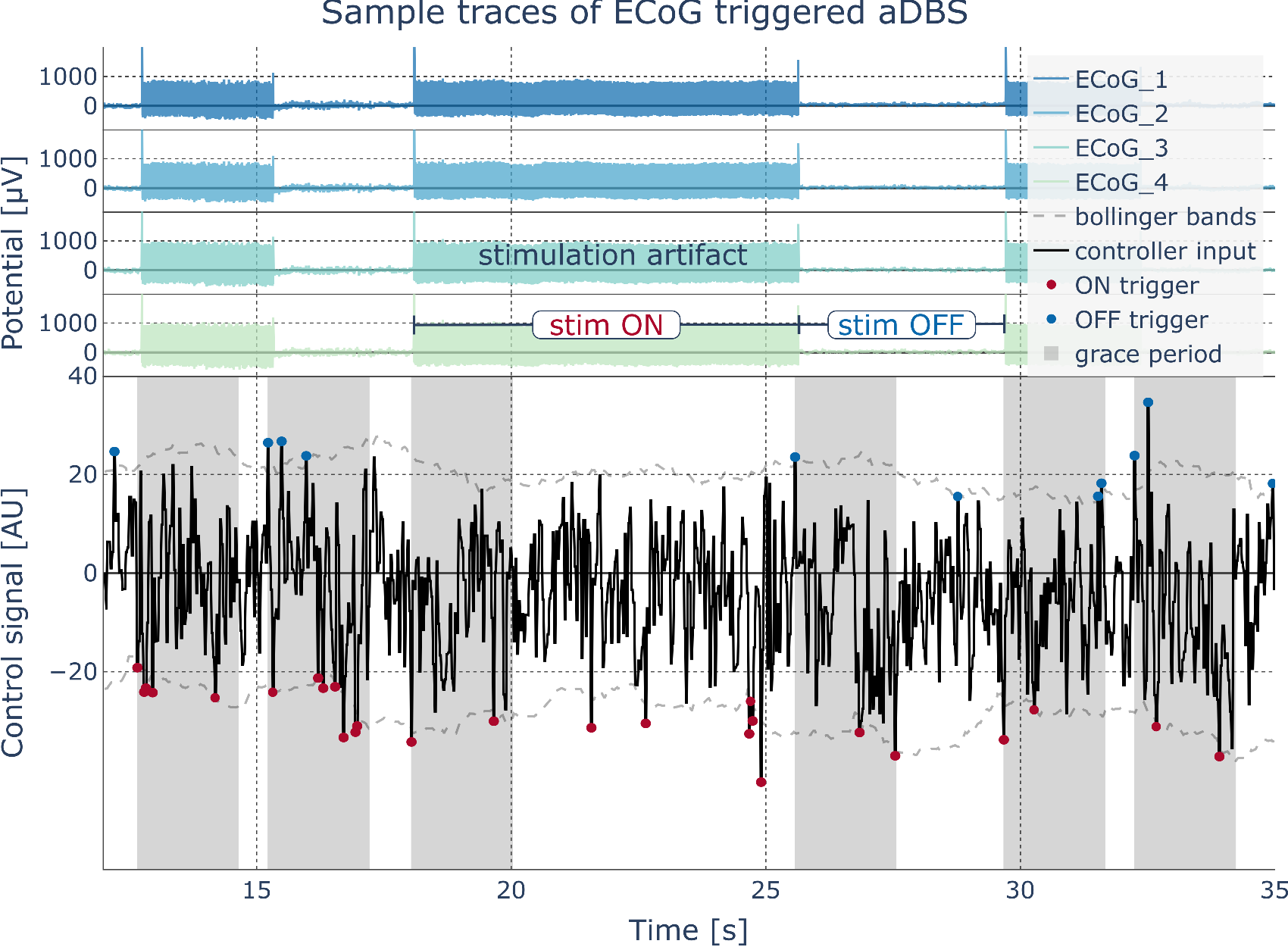}
  \caption{Example traces of the four ECoG channels and the signals used to control the closed-loop strategy. The first four rows from the top show the ECoG signals, clearly reflecting the stimulation artifacts in the stimulation ON segments. The last row shows the input to the Bollinger control module (continuous black trace), the Bollinger bands (dashed gray traces), trigger points (blue for stim OFF, red for stim ON) and grace periods (gray background) following a change of stimulation state. Note that the Bollinger band strategy was selected as it inherently produces crossings on either side of the Bollinger bands, which is an ideal test scenario. This is not a strategy to e.g., optimize the CopyDraw score, but would rather lead to stabilizing the scores to a medium level.}
  \label{fig:ClosedLoopExampleTraces}
\end{figure*}

% 11:30:54 to 11:53:44 for continuous aDBS
The main result of the closed-loop technical feasibility test on the patient with PD is that aDBS was successfully applied and tolerated without side effects by the patient for continuous windows of more than \qty{20}{min} (longest window on recording day 4, 11h30 to 11h53). Feasibility was tested during the execution of the CopyDraw task and with a decoding model based on the CopyDraw scores collected on day 3. Example traces of the control signal alongside ECoG raw recordings for the aDBS session on day 4 are shown in \autoref{fig:ClosedLoopExampleTraces}. Whenever the controller's input was crossing the Bollinger band limits, and if this crossing happened outside of the grace periods, a stimulation ON or OFF trigger was emitted. The stimulation change is visible by the stimulation artifacts appearing or disappearing in all 4 ECoG traces respectively.

\begin{figure*}
  \centering
\includegraphics[width=\textwidth]{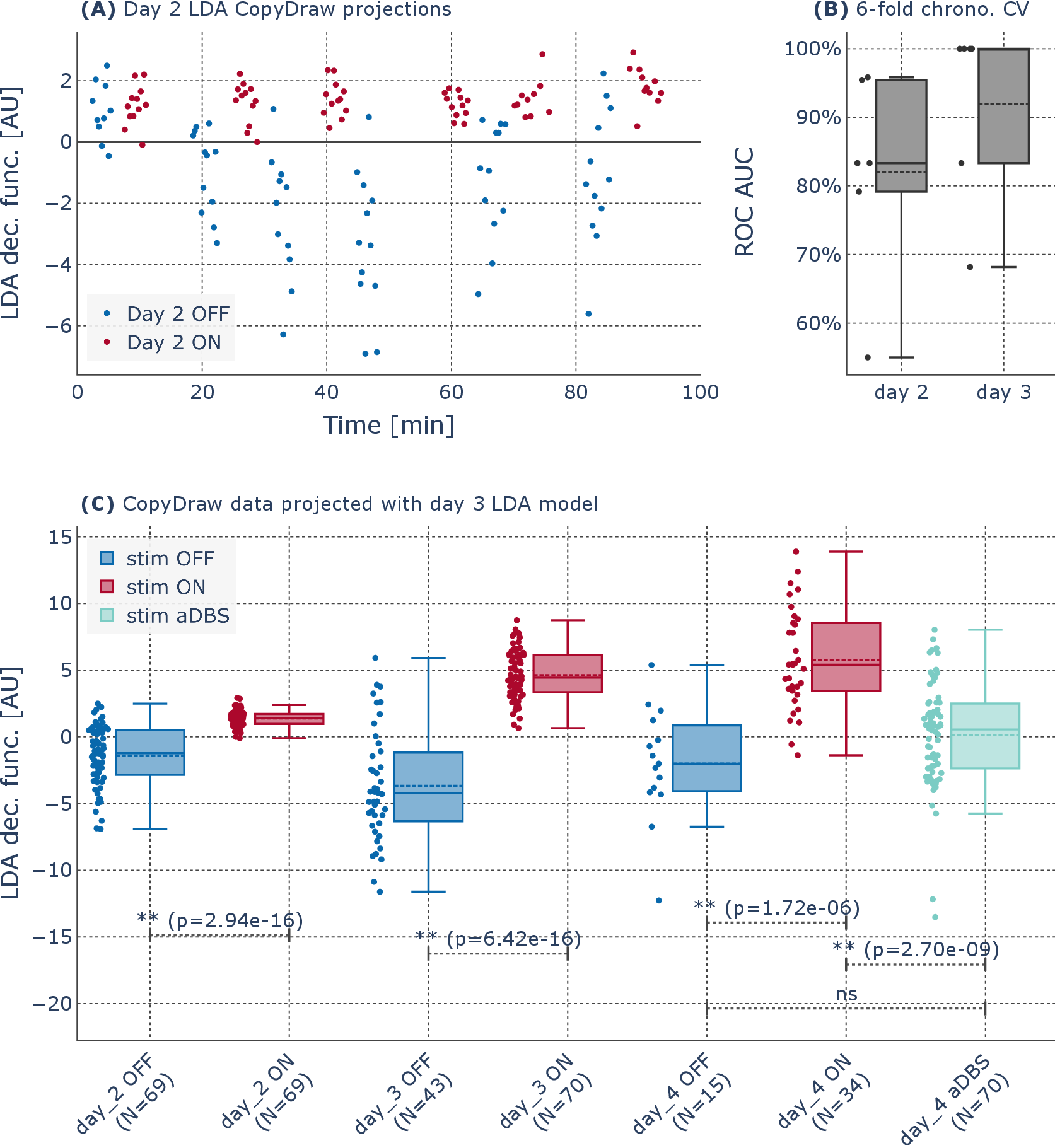}
  \caption{CopyDraw results: \textbf{(A)} CopyDraw scores, i.e., decision function values of a regularized LDA classifier, for individual trials of day 2. The y-axis shows the decision function values, the x-axis depicts the time since start of the first trial in minutes. Note that trials with pen lift-off were excluded from the analysis, leading to an unequal number of trials within individual blocks. This visualization serves to show the stability across time, but is generated from an LDA model fitted on the whole data of this measurement day. As a result, it cannot be used to interpret generalizability. \textbf{(B)} area under the receiver operating characteristic curve (ROC AUC) for decoding the stimulation state, estimated by a 6-fold cross-validation respecting chronological order, i.e., temporally neighboring ON and OFF blocks were considered as test folds. \textbf{(C)} overview of the CopyDraw scores from all three measurement days, predicted with the LDA model trained on day 3 data only. Significance indicators (**) mark test statistics of the Mann-Whitney U test with uncorrected p-values < 0.01, and \textit{ns} stands for a p-value > 0.05. The horizontal solid lines in the box plots (B) and (C) indicate the median values, while dashed lines show mean values.}
  \label{fig:CopyDrawDay3}
\end{figure*}

\begin{figure*}
  \centering
\includegraphics[width=\textwidth]{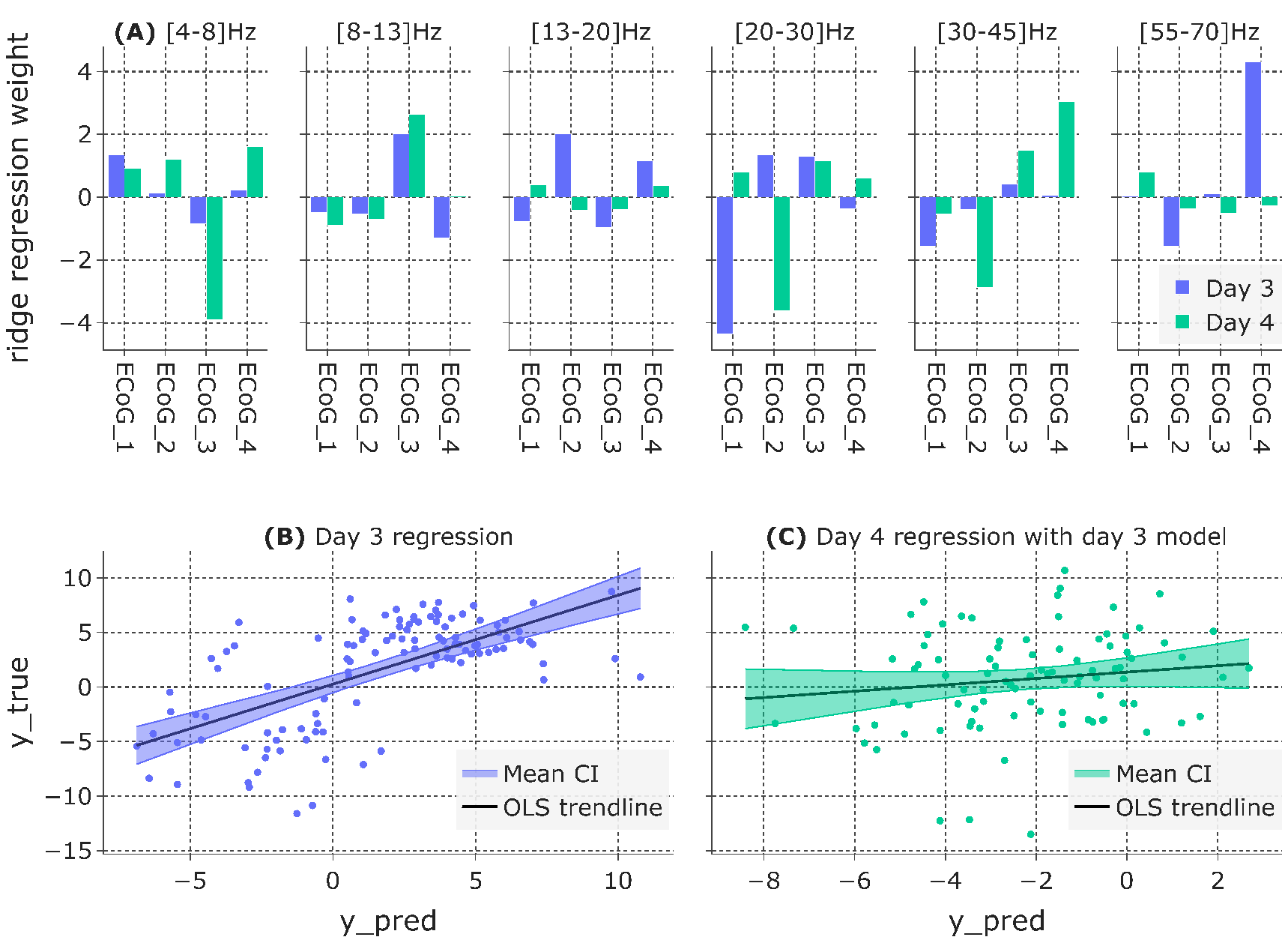}
  \caption{Investigation of the regression model: \textbf{(A)} shows coefficients of the ridge regression model predicting the CopyDraw scores of day 3 and day 4. All CopyDraw scores were derived from an LDA trained on day 3 data only. The models were trained on the full data of each of the days for visualizing the weights. \textbf{(B)} shows the day 3 ridge regression predicted vs actual CopyDraw scores. Predicted scores represent predictions collected during the 6-fold cross-validation. An ordinary least-squares (OLS) trend line was fitted with intercept and is shown with \qty{95}{\%} confidence intervals (CI). The slope is significant with P < 0.01 while the intercept is not significant P = 0.51. \textbf{(C)} is similar to (B) but shows a model trained on all day 3 data to predict CopyDraw scores for day 4. The slope is not significant for the OLS fit with P=0.17 while the intercept term is significant with P=0.04.}
  \label{fig:RidgeCoeffAndDecoding}
\end{figure*}

The patient showed a clear DBS-induced effect upon the CopyDraw performance, see \autoref{fig:CopyDrawDay3} (A) as an example for day 2.  \autoref{fig:CopyDrawDay3} (B) summarizes the stimulation-induced behavioral differences for the two sessions on day 2 and day 3. It shows the mean ROC AUC for a 6-fold chronological cross-validation of the LDA model, \qty{82}{\%} for day 2 and \qty{92}{\%} for day 3.
Please note that no separate evaluation could be conducted for the session on day 4, as the amount of data collected under stimulation OFF and ON conditions was insufficient for this day to fit a comparable LDA model with the same cross-validation procedures. 

CopyDraw scores were computed for trials of all three measurement days using the LDA model fitted on the CopyDraw data of day 3, see \autoref{fig:CopyDrawDay3} (C), to validate the stability of this label encoding across the measurement days. The resulting CopyDraw scores are significantly different (Mann-Whitney U test, P<0.01) between ON and OFF conditions for all measurement days. The CopyDraw scores obtained under the closed-loop condition (aDBS) are significantly different from the stim ON recordings, but not from stim OFF recordings obtained on day 4.

Although the decoding quality has not been the objective for this first technical feasibility experiment, a rudimentary analysis of the decoding model is presented in \autoref{fig:RidgeCoeffAndDecoding}. The model coefficients for two models fitted on CopyDraw trials of day 3 and trials of day 4 are shown in \autoref{fig:RidgeCoeffAndDecoding} (A). The scatter plots in \autoref{fig:RidgeCoeffAndDecoding} (B) and (C) show the CopyDraw scores predicted from ECoG data vs.~the (true) CopyDraw scores from the LDA decision function. For \autoref{fig:RidgeCoeffAndDecoding} (B), prediction results are taken from the 6-fold chronological CV. Mean Pearson's correlation across folds is $\rho=$\qty{71}{\%} (with a chance level of $\rho_{chance}=$\qty{21}{\%}). A linear fit using ordinary least-squares (OLS) regression shows a significant slope, P<0.01, but no significant intercept parameter (P=0.51). For \autoref{fig:RidgeCoeffAndDecoding} (C), a ridge regression model trained on all data from day 3 was used to produce the predicted scores. The Pearson's correlation $\rho=$\qty{14}{\%} for day 4 data is below chance level ($\rho_{chance}=$\qty{23}{\%}). An OLS fit shows no significant slope (P=0.17) while the intercept is significant (P=0.04).

\subsection{Other patient experiments conducted with Dareplane}
Including the two open-loop sessions and the single open- and closed-loop session (day 4) with the patient referred to above, a total of 19 open-loop DBS sessions were realized using the Dareplane platform. These sessions were conducted with 7 patients with externalized DBS leads, who received DBS treatment for PD or major depression disorder (MDD). See \autoref{tab:other_experiments} for an overview of recorded modalities - the presented patient is PD4.
All experiments were conducted with the same hardware, i.e., the same experiment notebook which was used for the benchtop evaluations. During the 19 sessions, the Neuro Omega system was integrated into Dareplane to record data and provide stimulation. It was controlled via the \dpm{dp-ao-communication} module.
During the recording sessions, we monitored the LSL signals on a separate notebook to reduce the CPU load of the main experiment notebook.
As the module orchestration via the \dpm{dp-control-room} uses TCP communication, additional hardware separation would be possible, e.g., by running the paradigm on a dedicated machine. This would however introduce the problem of drifting clocks.
A potential resolution would be to either restart or resynchronize the clocks for LSL frequently~\cite{Artoni2018} or, if possible, to use hardware markers that are integrated into the source signal directly. All recording sessions, with the exception of PD4, were purely offline recordings. Data from these sessions are not evaluated here as it would not provide proof of feasibility for the closed-loop processing capabilities.

\begin{table*}
    \centering
    \small
    \begin{tabular}{l|c|l|l}
    \toprule
    Patient & N$_{sessions}$ & \multicolumn{2}{c}{Modalities (channel count)} \\\cline{3-4}
            &                &  \qty{22}{kHz} sampling & \qty{5}{kHz} sampling     \\
    \midrule
    PD1 & 3 & LFP(16), ECoG(4), EOG(4) & HR(1), RR(1), GSR(1) \\
    PD2 & 3 & LFP(16), ECoG(4), EOG(4) & HR(1), RR(1), GSR(1), EEG(85) last session only\\
    PD3 & 3 & LFP(16), ECoG(4), EOG(4) & HR(1), RR(1), GSR(1)\\
    PD4 & 3 & LFP(16), ECoG(4), EOG(4) & HR(1), RR(1), GSR(1) - only in session 1 and 2\\
    MDD1 & 3 & LFP(16), ECoG(4), EOG(4)& HR(1), RR(1), GSR(1)\\
    MDD2 & 2 & LFP(16), ECoG(4), EOG(4)& HR(1), RR(1), GSR(1)\\
    MDD3 & 2 & LFP(16), ECoG(4), EOG(4)& HR(1), RR(1), GSR(1)\\
    \bottomrule
    \end{tabular}
      \caption{Overview of recordings conducted with Dareplane experimental setups at the UMC Freiburg. The presented CopyDraw evaluation and the closed-loop feasibility test refer to patient PD4. All other measurements were pure offline recordings. Recordings via the Neuro Omega were collected at \qty{22}{kHz} while recording via the BrainAmp ExG were collected at \qty{5}{kHz}. Heart rate (HR), respiratory rate (RR), and galvanic skin response (GSR) were collected using the according BrainAmp ExG sensors, while EEG was recorded using a BrainAmp with a gel based passive BrainCap (Brain Products, Gilching, Germany). HR, RR and GSR were not recorded during the closed-loop experiment to reduce the load on the experiment notebook.}
    \label{tab:other_experiments}
\end{table*}

\subsection{c-VEP experiment}
All three participants were able to spell the target sentence using the c-VEP interface. The performance metrics for all runs are shown in \autoref{tab:cvep_results}. Overall mean values across all subjects show an average correct symbols per minute (CSPM) rate of $13.247$, with an accuracy of \qty{90}{\%}, an average trial time of \qty{2.712}{s}, and an ITR of \qty{81.942}{bits/min}. Subject p002 shows the highest average performance with CSPM of 16.522, accuracy of \qty{93}{\%}, average trial time of \qty{2.133}{s}, and ITR of \qty{100.296}{bits/min}. Subject p003 has similar average performance metrics with CSPM of 13.247, accuracy of \qty{91}{\%}, average trial time of \qty{2.504}{s}, and ITR of \qty{86.044}{bits/min}. While the run (a) of subject p001 shows performance in line with the other subjects, the average performance of p001 is lower with CSPM of 9.227, accuracy of \qty{85}{\%}, average trial time of \qty{3.521}{s}, and ITR of \qty{59.485}{bits/min}.

\begin{table*}
\centering
\begin{tabular}{llrrrr}
\toprule
Subject & Run & CSPM & Accuracy & Average trial time [s] & ITR [bits/min] \\
\midrule
p001 & a & 11.105 & 91\% & 3.423 & 67.988 \\
p001 & b & 9.376 & 85\% & 3.412 & 60.402 \\
p001 & c & 7.198 & 78\% & 3.728 & 50.065 \\\cline{3-6}
$\text{mean}_{\text{p001}}$ & & 9.227 & 85\% & 3.521 & 59.485 \\
\midrule
p002 & a & 14.829 & 91\% & 2.268 & 91.149 \\
p002 & b & 17.291 & 94\% & 2.048 & 104.407 \\
p002 & c & 17.444 & 94\% & 2.021 & 105.332 \\\cline{3-6}
$\text{mean}_{\text{p002}}$ & & 16.522 & 93\% & 2.113 & 100.296 \\

\midrule
p003 & a & 15.089 & 94\% & 2.495 & 91.055 \\
p003 & b & 11.652 & 85\% & 2.547 & 75.137 \\
p003 & c & 15.240 & 94\% & 2.469 & 91.940 \\\cline{3-6}
$\text{mean}_{\text{p003}}$ & & 13.994 & 91\% & 2.504 & 86.044 \\
\midrule
\midrule
$\text{\textbf{mean}}_{\text{\textbf{all}}}$ & & 13.247 & 90\% & 2.712 & 81.942\\
\bottomrule

\end{tabular}
\caption{Overview c-VEP speller experiments. The information transfer rate (ITR) is including a \qty{1}{s} inter-trial-time. CSPM stands for correct symbols per minute.}
\label{tab:cvep_results}

\end{table*}

% %
\section{Discussion} % (fold)
\label{sec:discussion}

\subsection{The Dareplane platform}

In this work, we have presented Dareplane, a modular open-source platform for BCI experiments. 
By pushing the abstraction of the processing steps which are relevant for aDBS research, we have designed the Dareplane platform to stay modular and technology-agnostic. This was realized by using the well-established and -tested LSL~\cite{Kothe2024, LSL2024,Schulte2022,Artoni2018,Chuang2021}.
Our design resulted in a loose coupling of the modules and overall low requirements on the communication protocol.
It is therefore easy to integrate existing software into the platform, e.g., by providing a lean wrapper that processes PCOMMs and interacts with the LSL interfaces.
Although there is no preference for functional programming or object-oriented (OO) develop\-ment philosophies, the Dareplane platform does incorporate the \textbf{SOLID} principles~\cite{Martin2000} known from OO. They are, however, to be understood on a modular instead of a per-object level.
\textbf{S}ingle responsibility is realized by defining single modules for each processing step.
The \textbf{O}pen-closed principle is realized, as the platform is open for extensions by adding new modules, and closed to modification, as these new modules can be added without the need to change existing modules.
\textbf{L}iskov substitution can be understood as replacing modules for a given task in an existing setup, e.g., by replacing the behavior paradigm or the algorithm used for decoding.
\textbf{I}nterface segregation is inherent, as each module only provides PCOMM functionality specific to itself, and only queries LSL streams specified for its own use.
Lastly, \textbf{D}ependency inversion, which is also understood as depending on abstractions instead of concretions, was the core idea behind Dareplane's design. 

The above-mentioned features were achie\-ved at the expense of 1) a network communication overhead and 2) a memory overhead induced by every module having its own data buffers, and 3) the platform targeting a tech-savvy user group. Both, 1) and 2) are caused by the independence of each module and the data streaming via network communication. This is mainly a constraint to consider during the design process of a setup, and users should be aware of how much data is streamed, which usually is not a problem in the BCI context, with the potential exception of high frequency video data~\cite{Kothe2024}, or invasive single-unit activity recordings with high sampling rates and large channels counts. The same holds for the buffer design, which is a trade-off between the necessary data history and the available memory. However, this does not affect the permanent data recording, as raw signals and LSL streams can be written to hard-drive in parallel.
Furthermore, each module will be spawned in a separate subprocess. Depending on the module, it will include a kind of \code{main loop}, presenting a continuous CPU load which might exhaust the hardware capabilities. As the benchtop tests reveal, it is still possible to support high throughput with modest consumer-grade hardware, as opposed to other similar solutions which are further optimised for performance~\cite{Ali2024}, but require large amounts of memory or CPU power to achieve this performance.

Although Dareplane targets a technically versed user group, the platform is still very accessible, provided a user is proficient with the programming language they chose to implement their own modules with. Reusing existing modules based on, e.g., Python, requires the user to understand how to set up and manage a Python programming environment. 

While the current Dareplane modules use LSL~\cite{Kothe2024} for data streaming, there is no strict dependency on it. Modules could be implemented with other streaming technology or even allow for configurable streaming backends.
Compared to mature and feature-rich platforms like BCI2000~\cite{Schalk2004}, Dareplane provides less functionality out of the box in its current state. With its design of minimal requirements, implementing new functionality is made very accessible. As an example, working purely with interpreted software is possible, while compiled code can be used wherever needed for communication with APIs or when performance-critical components need to be sped up. Furthermore, Dareplane runs on UNIX and Windows systems alike, while platforms like BCI2000 can face functional constraints when being compiled for non-Windows systems. Cross-platform availability is of course easier for younger platform projects. Supporting Windows only might not be a downside, though, especially when targeting a user group that is not developing custom software for their experiments.

\subsection{Benchtop tests}

\paragraph{Observed latencies} - The latency and performance benchtop tests conducted with modest hardware have revealed the general possibility of processing data at high sampling rates, exceeding \qty{1}{kHz} in the Dareplane platform. Regarding latencies, we have identified the comparably slow responses of the stimulation hardware based on API triggers as the key bottlenecks for a closed-loop system.
Other platforms have considered a closed-loop performance as feasible, if round-trip times of less than \qty{15}{ms}~\cite{Ali2024, Ciliberti2017} can be achieved. These thresholds were considered for pure on-platform processing or by recording a digital output signal, but did not consider a delay by a potential feedback device, such as a neurostimulator or the refresh of a screen. Our benchtop test showed that, if omitting the stimulation device's response for a fair comparison, the Arduino ($\Delta_{q99}$SD + $\Delta_{q99}$DC + $\Delta_{q99}$CT = \qty{2.059}{ms}) and BIC-EvalKit ($\Delta_{q99}$SD + $\Delta_{q99}$DC + $\Delta_{q99}$CT = \qty{4.618}{ms}) tests stay below this value for the sum of their \qty{99}{\%} quantiles. The processing time for the Neuro Omega can be slightly above this value ($\Delta_{q99}$SD + $\Delta_{q99}$DC + $\Delta_{q99}$CT = \qty{17.941}{ms}), which is mainly caused by the pass-through time of data that is chunked in different clusters, which is clearly visible by the $\Delta$SD>\qty{10}{ms} values of \autoref{fig:BenchtopPerf} (D). This leads to $\Delta_{q99}$SD values around \qty{16}{ms}.

The results for the simplified Arduino test lead to a question as the observer (Arduino mean GPIO reaction $\Delta$TS) is much slower than observed GPIO response times of less than \qty{1}{ms}~\cite{Appelhoff2021} which was observed with a comparable LSL setup. A direct comparison, however, is not possible as the Arduino used in the experiment by Appelhoff et al.~\cite{Appelhoff2021} was programmed differently, also using a higher baud rate for the serial communication. Still, a systematic delay in the timestamps received from the oscilloscope module or API could be another explanation for the observer difference. A separate calibration procedure with the oscilloscope module will be necessary to rule out a systematic delay and will be part of future investigations.

The delay observed between crossing the signal threshold and the appearance of the stimulation artifact 20-\qty{40}ms later for the BIC-EvalKit and Neuro Omega is in line with delays reported by other groups stimulating on externalized leads~\cite{Little2013, Tinkhauser2017b}.

Whether this delay can be considered a sufficiently low latency depends on the use case. Given the long-tailed distribution and the occasional outliers with very long timestamp differences, it is clear that a Dareplane setup cannot provide tight timing guarantees running on a regular operating system. 
Dareplane provides only "soft realtime"~\cite{Patel2017} capability, which means that there is no guarantee for a processing step to be executed within a given time window. This characteristic is shared by other platforms running on regular operating systems. Only dedicated platforms for hard realtime processing are capable to provide guaranteed processing~\cite{Patel2017}. Considering the processing windows of current aDBS applications, which usually comprise a processing every few 10-\qty{100}{ms}~\cite{Little2013, Gilron2021, Swann2018} or even adjusting stimulation only once per second~\cite{Rosa2015}, the latencies achieved with Dareplane seem to be sufficient to realize similar protocols nonetheless.  

\paragraph{Time-critical markers for aDBS} - Looking forward, a recently proposed approach to aDBS is based on beta-burst predictive triggering~\cite{Sargezeh2024}. The authors used a neural-network classifier to predict the occurrence of beta bursts~\cite{Tinkhauser2017}, for which a reliable classification was achieved between 30-\qty{80}{ms} prior to burst onset~\cite{Sargezeh2024}. A response within this time window would be possible using the tested hardware.

Another marker discussed for aDBS of patients with PD is phase-dependent stimulation~\cite{Holt2019}, which is common in essential tremor where the frequency used as a biomarker typically is in the slower theta (5-\qty{8}Hz) range~\cite{Cagnan2017}. Furthermore, the neural oscillation is tightly coupled to the tremor frequency, allowing, e.g., an aDBS system based on accelerometric data~\cite{Castano2020b}.
Using the phase of a narrow-band oscillation in the context of PD as a marker for aDBS is to our knowledge only applied in animal models~\cite{McNamara2022} and discussed with simulation models of the thalamo-cortical system~\cite{West2022}.
Triggering stimulation phase-dependent on the beta oscillations is an interesting and technically challenging potential future use case for the Dareplane platform, which should be feasible given the observed latencies.

\paragraph{Timing uncertainty} - It is informative to also compare the measured latencies to signal-specific timing metrics to judge if latencies induced by the modular Dareplane structure are acceptable. For LSL streams of different hardware sources, timing can be analysed regarding fixed offsets, drifts and jitter. Fixed offsets are accounted for by LSL and are usually in the range of \qty{10}{ms}~\cite{Kothe2024}. Drifts caused by different hardware clocks are observed to be around \qty{0.03}{ms/s}~\cite{Artoni2018, Schulte2022} and can have a severe impact on decodability~\cite{Artoni2018}. Finally, jitter is observed to be around \qty{1.7}{ms}~\cite{Schulte2022, Artoni2018}. For BCI systems based on evoked responses, jitter values of \qty{10}{ms} and larger lead to a significant SNR reduction~\cite{Iwama2023}. Given these independent timing inaccuracies, we consider the overhead introduced by the modular structure of the Dareplane platform acceptable.

Data arriving in chunks provides another timing uncertainty that may compromise closed-loop response times.
Receiving data in chunks rather than on a per-sample basis is hardware-specific and is also to be expected for other signal recording hardware~\cite{Siegle2017} including IPGs~\cite{Gilron2021}.

\paragraph{Load restrictions} - While testing the benchtop setups, the standard Windows system monitor showed total CPU usages larger than \qty{90}{\%} when the modules were configured with higher sampling rates, e.g., processing modules operating at \qty{1}{kHz} on data from the Neuro Omega which was streamed at \qty{22}{kHz} for 24 channels.
We empirically observed data loss whenever the system was above a CPU load of \qty{95}{\%} for more than a few seconds. This implies that whenever a demanding decoding method is considered, either a slower sampling rate (within the modules) needs to be selected, or more potent hardware should be used. Given the modular nature of the setup, which spawns separate processes for each module, we expect robust performance increases by the kill-it-with-iron approach of facilitating more cores.

\paragraph{Comparability of the setups} - Comparing the three benchtop tests it is interesting to note that the time delay between the LSL timestamp for triggering a stimulation and the arrival of the stimulation artifact of the Arduino Uno GPIO $\Delta^{\text{AUno}}_{TS}=$\qty{12}{ms} is at least roughly comparable to the latencies achieved with the Neuro Omega system $\Delta^{NO}_{TS}=$\qty{46}{ms} or CorTec's BIC-EvalKit $\Delta^{BIC}_{TS}=$\qty{22}{ms}. This suggests that realistic benchtop experiments can be created with low-cost, accessible hardware, which allows to pre-test setups when access to the clinical hardware is limited, or would allow to realize student projects under financial constraints.

\subsection{Patient experiments}

Regarding the feasibility of Dareplane to perform an aDBS experiment with a patient in the loop, we can conclude that it was technically feasible to use the Dareplane platform. The achieved update cycles with up to \qty{60}{Hz} are comparable to update speeds in current aDBS experiments, ranging from ~\qty{2}{Hz}~\cite{Swann2018} to a few \qty{10}{Hz}~\cite{Gilron2021}. The resulting stimulation pattern was tolerated well by the patient for extended stimulation periods, which were similar in length to the time periods reported by other studies with stimulation on externalized leads~\cite{Velisar2019}. 

The chosen control strategy aimed to ensure that a switching of the stimulation was triggered in order to run a technical feasibility test of the closed-loop functionality. We did not aim to find the optimal control strategy that would result in the best CopyDraw performance, or would even lead to a clinically optimal symptom improvement. The resulting CopyDraw scores observed on day 4 during the aDBS session are not significantly different from stimulation OFF blocks of the same day. It is to be noted that the CopyDraw score by design leads to an optimized discrimination between stimulation ON and OFF. It does not describe the quality of the copied trace as such. Therefore it is possible that an objective rater would judge the quality of the CopyDraw traces produced by the patient during stimulation OFF state as higher than traces produced during the ON state. Overall, the stimulation effect on the CopyDraw score was stable across all three measurement days,  as can be seen by the significant difference of the stimulation ON vs.~OFF blocks from just one LDA model (see~\autoref{fig:CopyDrawDay3}). Observing such a stable behavioral effect, paired with the high ROC AUC scores of the CopyDraw LDA models for day 2 (\qty{82}{\%}) and day 3 (\qty{92}{\%}) is not common for patients with PD performing CopyDraw~\cite{Castano2020}. The labels therefore provided high-contrast information for the neural decoding. This is in line with the high correlation values achieved by the ridge regression used with the ECoG multi-band features of day 3. 
%The observed feature weights for day 3, see~\autoref{fig:RidgeCoeffAndDecoding} (A), are in line with published findings that a reduction in beta power and an increase in gamma power over M1 are associated with the effect of DBS~\cite{Muthuraman2020}. 
Contrary to the behavioral decoding, the transfer of the ridge regression model for neural decoding did not generalize to the different day 4 -- a significant correlation could not be achieved between the predicted and the actual CopyDraw scores. This potentially indicates that neural markers for the CopyDraw performance are unstable over days, which is also reflected in the very different ridge regression weight distributions between both days. It is to be noted however, that these regression weights are difficult to interpret, as a large weight could either be related to a strong signal or to a noise component that is to be projected out~\cite{Haufe2014}. Furthermore, while unstable markers indicate the need for session-specific marker identification known from EEG data~\cite{Castano2020}, most trials (N=70, \autoref{fig:CopyDrawDay3}(C)) on day 4 were conducted under an aDBS condition which had not been observed by the model on the previous days. The impact of such a changed stimulation protocol upon the decoding performance of CopyDraw is unknown and needs to be subject to further investigation. Explicitly, no LDA model for day 4 could be trained as there was an insufficient number of trials under pure stimulation ON and OFF collected, in favor of performing more blocks under aDBS. Thus no ROC AUC for the CopyDraw was reported for day 4. Another limitation of the presented proof of concept is that the decoding approach was not exhaustively investigated. Including band-power features beyond \qty{70}{Hz} or a different binning of the bands could improve the decoding performance. Furthermore, the used moving-average approximation for the band power is only a cost-efficient, but potentially crude approximation which could be optimized by considering the variance of the (narrow) band-filtered signal, or a Hilbert transform in a causal filter design for allowing comparability to online decoding. Additionally, a different choice for ground and reference, e.g., using one of the ECoG channels, should be considered to reduce noise and correlation in the signal.

\subsection{c-VEP experiments}
The performance metrics of all three participants with mean ITR of \qty{81.942}{bits/min}, are similar to the published results for the rCCA decoding approach~\cite{Thielen2015} with early stopping~\cite{Thielen2021}. While using a different keyboard grid with only n=29 characters, Thielen et al.~\cite{Thielen2021}, report an average ITR of \qty{67.9}{bits/min} (max. 106.7, min. 44.2) for the copy-spelling task. With this successful replication of the c-VEP speller on three healthy participants, we demonstrate the use case of Dareplane for a non-invasive BCI. With the provided setup script, we hope to foster adaptation of the Dareplane platform, as replicating the experiment requires a setup available in many labs: an LSL integrated EEG system, a screen for stimuli presentation, and a PC for running the software.

\subsection{Future development}
The presented Dareplane setups were all hosted on a single machine. Given Dareplane's general capability of running on distributed systems, an interesting next step would be to consider a setup involving at least two machines with simultaneous processing tasks. While performance measurements exist for the use of multiple machines for data acquisition and processing~\cite{Iwama2023, Dasenbrock2022}, it is not clear how distributing decoding and/or control strategies tasks over different machines would impact the overall latency. The results of the performance benchmark presented in this paper are to be understood as an out-of-the box potential with modest hardware. A more comprehensive benchmarking could consider more potent hardware, or the optimization of individual modules, e.g., by aligning the update cycles of the main loops of individual modules. Finally, an evaluation using a network stack in combination with a realtime operating system could provide interesting insights into how far the modularity in combination with LSL for data communication can be pushed.

\section{Conclusion} 
\label{sec:conclusion}

While the \textit{decoding} of biomarkers for closed-loop neurotechnology systems like BCIs and aDBS is a very active research topic, the problem of \textit{optimizing} adaptive stimulation seems to be harder and so far is addressed by few publications only. The reason for this may partially be explained by a lack of a versatile aDBS software platform. With Dareplane, we have developed and offered an open-source modular software platform to the community for running BCI- and specifically aDBS experiments. The platform uses TCP communication for exposing and triggering functionality and uses LSL for sharing and recording of data. Dareplane's tech stack is mostly technology-agnostic and allows integrating existing code with very little overhead. The conducted benchmarks show a sufficient latency for most BCI- and aDBS tasks. We tested the technical feasibility for aDBS in a clinical experiment with a single patient with PD and demonstrated its general applicability in a non-invasive c-VEP BCI application. As an open-source project, we hope to steadily grow the platform and improve quality and functionality with a wider range of developers and contributors in the long run.

\ack
MD receives funding from the Dareplane collaboration project, which is co-funded by PPP Allowance awarded by Health~Holland, Top Sector Life Sciences \& Health, to stimulate public-private partnerships as well as by a contribution from the Dutch Brain Foundation. The Dareplane project also received in kind contributions by CorTec (Freiburg, Germany) and Newronika (Milan, Italy). JP is funded by the Deutsche Forschungsgemeinschaft (DFG) under the Walter Benjamin Program (Project number 510112977). The PD-Interaktiv I study is funded by the Bundesministerium für Bildung und Forschung (Grant 16SV8011). BS receives a research grant from Ceregate (Hamburg, Germany) and honoraria as an advisor for Precisis (Heidelberg, Germany), both unrelated to this work. VC receives a collaborative grant from BrainLab (Munich, Germany). He serves as an advisor for Aleva (Lausanne, Switzerland), Ceregate (Hamburg, Germany), Cortec (Freiburg, Germany) and Inbrain (Barcelona, Spain). He has an ongoing IIT with Boston Scientific (USA). He has received travel support and honoraria for lectures from Boston Scientific (USA), UNEEG Medical (Munich, Germany), and Precisis (Heidelberg, Germany).

\section*{Author contributions}
\textbf{MD}: Platform design and development, conducting patient experiments, implementation c-VEP speller, writing original draft.
\textbf{JP}: Platform design and development, open-loop patient experiments.
\textbf{JT}: implementation c-VEP speller, conducting c-VEP speller experiments.
\textbf{MT} and \textbf{MJ}: Platform design.
\textbf{BS} and \textbf{VC}: Patient recruitment, surgery, and safety surveillance during experiments.
\textbf{All authors:} Writing - review \& editing.

\section*{References}

\bibliographystyle{vancouver}   

\bibliography{bibliography}

\end{document}